\documentclass[11pt,fancychapters]{report}
\usepackage[a4paper, total={6in, 8in}]{geometry}
\usepackage{listings}
\usepackage{color}
\usepackage{xcolor}
\usepackage{setspace}
\usepackage{acro}
\usepackage{amsmath}
\usepackage{amsthm}
\usepackage{graphicx}
\usepackage{geometry}
\usepackage{subcaption}
\usepackage{cancel}
\usepackage{tikz}
\usepackage{color}
\usepackage{dsfont}
\usepackage[frozencache,cachedir=.]{minted}
\usepackage{caption}
\usepackage{comment}
\usepackage{glossaries}
\usepackage{authblk}
\usepackage[backend=bibtex]{biblatex}
\usepackage{doi}
\usepackage{colortbl}
\usepackage{booktabs}
\usepackage{hyperref}
\usepackage{multirow}

\usepackage{pythontex}
\usepackage{listings}

\usepackage{multicol}

\definecolor{codegreen}{rgb}{0,0.6,0}
\definecolor{codegray}{rgb}{0.5,0.5,0.5}
\definecolor{codepurple}{rgb}{0.58,0,0.82}
\definecolor{backcolorcream}{rgb}{0.95,0.95,0.92}

\lstdefinestyle{solutionstyle}{
    backgroundcolor=\color{backcolorcream},   
    commentstyle=\color{codegreen},
    keywordstyle=\color{magenta},
    numberstyle=\tiny\color{codegray},
    stringstyle=\color{codepurple},
    basicstyle=\ttfamily\footnotesize,
    breakatwhitespace=false,         
    breaklines=true,                 
    captionpos=b,                    
    keepspaces=true,                 
    numbers=left,                    
    numbersep=5pt,                  
    showspaces=false,                
    showstringspaces=false,
    showtabs=false,                  
    tabsize=2
}

\lstdefinestyle{examplestyle}{
    commentstyle=\color{codegreen},
    keywordstyle=\color{magenta},
    numberstyle=\tiny\color{codegray},
    stringstyle=\color{codepurple},
    basicstyle=\ttfamily\footnotesize,
    breakatwhitespace=false,         
    breaklines=true,                 
    captionpos=b,                    
    keepspaces=true,                 
    numbers=left,                    
    numbersep=5pt,                  
    showspaces=false,                
    showstringspaces=false,
    showtabs=false,                  
    tabsize=2
}

\newcounter{example}[section]
\newenvironment{example}[1][]{\refstepcounter{example}\par\medskip
   \noindent \textbf{Example~\theexample. #1} \rmfamily}{\medskip}

\newtheorem{theorem}{Theorem}[section]   
\newtheorem{exercise}[theorem]{Exercise}
\newtheorem{solution}[theorem]{Solution}
\theoremstyle{definition}
\newtheorem{definition}{Definition}[section]

\usetikzlibrary{calc,trees,positioning,arrows,chains,shapes.geometric,%
    decorations.pathreplacing,decorations.pathmorphing,shapes,%
    matrix,shapes.symbols,tikzmark}
\hypersetup{
    colorlinks,
    citecolor=black,
    filecolor=black,
    linkcolor=black,
    urlcolor=black
}

\definecolor{DarkerGreen}{RGB}{0,179,45}

\newtheorem{exmp}{Example}[section]

\newcommand{\splitatcommas}[1]{\begingroup\lccode`~=`, \lowercase{\endgroup
    \edef~{\mathchar\the\mathcode`, \penalty0 \noexpand\hspace{0pt plus 1em}}%
  }\mathcode`,="8000 #1%
  }

\tikzset{
>=stealth',
  punktchain/.style={
    rectangle, 
    rounded corners, 
    draw=black, very thick,
    text width=10em, 
    minimum height=3em, 
    text centered, 
    on chain},
  line/.style={draw, thick, <-},
  element/.style={
    tape,
    top color=white,
    bottom color=blue!50!black!60!,
    minimum width=8em,
    draw=blue!40!black!90, very thick,
    text width=10em, 
    minimum height=3.5em, 
    text centered, 
    on chain},
  every join/.style={->, thick,shorten >=1pt},
  decoration={brace},
  tuborg/.style={decorate},
  tubnode/.style={midway, right=2pt},
}

\newglossaryentry{non parametric}
{
    name=non parametric,
    description={Non parametric statistics does not assume that the underlying probability distribution has a predefined form, e.g., of type normal or exponential, and nevertheless is able to apply statistical inference to the problem. }
}

\newglossaryentry{control interval}
{
    name=control interval,
    description={For a random variable $X$, a control interval is a segment $[low, high]$ such that the probability of having a random variable materialising outside the segment is low}
}

\newglossaryentry{ML model}
{
    name=ML model,
    description={Depending on the learning task this may stand for different things.  In general a ML model some determining function that has predictive capability.   For example, in the case of a clarification ML task it will be a function that given a new objected, e.g., en image, 
determine its type, e.g., whether it is a cat or a dog. }
}

\newglossaryentry{bootstrapping}
{
    name=bootstrapping,
    description={Given a sample, $P$, from some distribution $F$, the empirical distribution of the sample $S$ represents the distribution, $P$, if the sample is big enough.  We can thus re-sample from $S$ with replacement to obtain fresh samples that approximate fresh samples from $P$.  This process is called bootstrapping.   }
}

\bibliography{main}
\begin{document}

\title{
Effective Technical Reviews
}

% Comment this next line out if we need to hide our working sections
%\def\SHOWWORK{}

\author[1]{Scott Ballentine}
\author[2]{Eitan Farchi}
\affil[1]{IBM Systems, sballen@us.ibm.com}
\affil[2]{IBM Haifa Research Lab, farchi@il.ibm.com}
\date{\today}

\maketitle
\pagenumbering{gobble}
\newpage
\pagenumbering{roman}
\setcounter{tocdepth}{3}
\tableofcontents
\begin{abstract}
There are two ways to check if a program is correct, namely execute it or review it.  While executing a program is the ultimate test for its correctness reviewing the program can occur earlier in its development and find problems if done effectively.  This work focuses on review techniques.  It enables the programmer to effectively review a program and find a range of problems from concurrency to interface issues.  The review techniques can be applied in a time constrained industrial development context and are enhanced by knowledge on programming pitfalls. 
\end{abstract}
\newpage
\pagenumbering{arabic}

\chapter{Introduction and background}

%TBC - general - discuss when to have a review meeting and when not to have one.  What are the tradeoff.  The benefit of undertanding and what can be done effectivally without undertanding.  

Why are we conducing technical reviews?  Many have the following experience: when writing a paper or an essay there is a point in which you find yourself staring at a page and nothing comes to you mind.  You are no longer able to improve your write-up!  That is when another, less involved person, may help.  The other person will bring in a new perspective, a new point of view.   The other person as an outsider will point to things you are missing.   If the other person is an expert, all the better!   In any case, this outside view at the point in which you are no longer able to improve your work by yourself is the reason technical reviews were introduced into the software development process. 

Formal technical reviews were introduced by Fagan in the 1970's \cite{fagan}.  They first focused on code reviews but quickly expended to the review process of any artifact created as part of the software development process, namely, requirements documents, design documents, test plans and customer documentation.  Back then, reviews included a preparation phase and a meeting.  People would read the artifact being reviewed ahead of the review meeting and come prepared.  In the review meeting issues where raised utilizing to preparation, they lead to brief discussion and additional issues.  Statistics showed high effectiveness of technical reviews.  Today although software engineering development processes changed dramatically and review meetings are not in fashion the effectiveness of technical reviews is still generally recognized.  One of the challenges of traditional review process was their overhead.   To sum up, they were highly effective but slow and largely connected with a software development project that included a lot of documentation.

When technical reviews were first introduced, most software development followed the waterfall model.  In this model, there are several phases of development, and each phase must be completed, with the appropriate approvals, before moving on to the next phase \cite{royce}.  This means that coding cannot start until the design phase has completed, and testing cannot start until the coding phase has completed.  Despite their overhead, Fagan-style inspections fit very well into this model of software development for two reasons.  First, having technical reviews were part of the required approval to move on to the next development phase.  Second, waterfall development projects took a long time to complete, typically measured in months or years, so the entire development model had a lot of overhead already.  Compared to the overall water fall development process, the overhead introduced by Fagan reviews was negligible and the benefits are clear cut. 

Software development has evolved significantly since then.  Most development projects now utilize some form of Agile development.  There is a preference toward regularly providing small amounts of working code that build upon each other, rather than delivering large amounts of code at the end of a very large project.  Typically, agile teams work in small, time-boxed iterations, and deliver code on a regular cadence that can be measured in weeks.  Also, related methodologies such as test-driven development allow for almost instant testing of software, providing the ability to fix problems quickly.  With the speed of modern development, a traditional review process that requires a lot of overhead has fallen out of favor.

This work introduces an in-depth view of technical reviews.  We are shooting for review techniques that can be done effectively in software projects that a market driven, time constrained and not necessarily heavy on documentation.  Thus, the techniques can be applied in an Agile development environment but are in fact largely independent of the specific software engineering development methodology that is being applied. 

Review meetings made a lot of sense when development teams were located in one site.  As teams become globalized and the workplace more digital it is no longer make sense to strive for a F2F review process.  With the advance of remote communication capabilities, review meeting and offline reviews can be highly effective if properly done.  We also address this topic in the book.

Given that most software projects are time constrained, it is necessary to make the most of what little review time a development team can afford.  A little bit of preparation can go a long way.  Providing some reference material, such as design documentation, sample output, or processing flows, can help move the review process along quickly and help focus it on design issues that are indeed more costly.  Also, rather than reviewing the entire artifact, if there are specific concerns identified in advance, the review can be focused on those concerns, which can greatly reduce the amount of time spent reviewing.

Depending on the concerns with the artifact, certain review techniques might make more sense than others.  Some of the techniques we cover in this book might be considered "common sense", but that does not make them any less valuable.  One example of a common sense technique is having a checklist of common problems or errors, based on prior experience, to look for when reviewing the artifact.  Checklists, when done properly, can be extremely effective in uncovering issues.  One should avoid the pitfall of a large checklist.  A checklist of a hundred items will not be used.  Customized six to ten items that focus on the review concerns can be helpful and effective.

Another technique we discuss is the idea of an obligation or a contract.  Every programming interface in a software system, whether it be a low-level method or function internal to the product, or an external product interface, has a set of expectations it must adhere to, as well as a set of expectations its caller must adhere to.  These expectations can be described as a contract between the interface and its caller, or an obligation of the caller or an obligation of the interface.  Focusing on these obligations is useful when defining a new interface or invoking an existing interface in a new context.

\begin{exercise}
\label{openandwrite}
What is wrong with the following code snippet, if anything:  
\begin{lstlisting}[language=C]
$fd = open("myFile", w);
buf = read(fd);
\end{lstlisting}

Choose the best possible answer.
\begin{enumerate}
\item Nothing is wrong with the code snippet.
\item The memory for buffer $buf$ is not allocated.
\item Indeed we would like to check if the buffer memory is allocated or not, but clearly we are not checking if the open operation was successful.
\item Probably the buffer is not allocated.  It is OK that we are not checking the result of the open operation as it is unlikely that the open operation fails.
\end{enumerate}
Solution: \ref{openandwrite_solution}
\end{exercise}

Desk checking is an old technique, where you effectively "play computer", walking through a segment of code by hand based on a specific set of input values.  It might seem that this is not a very effective exercise, since you could simply test the code, but reviewing this way can often expose errors and does not require to setup a complicated environment to run the test.  Although a specific input might work successfully when testing, reviewing the code using this method might show some internal logic errors or expose that a slight change to the set of input values would give an incorrect result.  We will also discuss test selection, the process of choosing a subset of tests to review or execute, because most programs have a large number of inputs and it would be impossible to review using every possible combination of input values.

The desk checking technique works well with a simple program that runs sequentially.  Most modern software is not simple, as it often has multiple threads or processes that interact with each other.  A variation of the desk checking technique, called the interleaving review technique (IRT), can be applied to these processes to uncover interaction or synchronization/serialization problems.

%TBC 
%\begin{itemize}
%    \item Pros and Cons
%    \item Compare reviews vs. testing
%\end{itemize}

\chapter{Review mindset and pitfalls}

No matter how much we streamline the review process, it still takes time and resources to perform a review.  If you are going to invest that time and energy, it needs to provide value.  There are many ways a review can go wrong and waste everybody's time.

The first question to ask yourself is "Why are we holding a review"?  In many cases, reviews are required because of business or regulatory requirements.  If reviews are only held to "check the box" so that the developer can merge his or her pull request and move on to the next work item in the queue, they might provide some minimal value, but they are not likely to provide a lot of benefit to the organization in the long run.

It is much better to go into a review with the goal of finding problems.  Even the most experienced and skilled developer will inject bugs into their code.  That is different than proving that there are no errors.  For all but the most simple pieces of code, it is impossible to prove that errors do not exist.  We can only prove that we did not find any errors.

Every review participant plays a role in the review.  For the purposes of a review, we will refer to the developer who wrote a piece of code as the code developer. Everybody in the review acts as a reviewer, including the code developer.  Any programmer who has had the experience of having to explain a piece of code to another person, or even to an inanimate object such as a rubber duck\cite{pragmaticprogrammer}, can understand how the code developer implicitly becomes a reviewer of their own work.  There are other roles that can be played in the review, some of which are specific to certain review techniques, which we will cover later.

This leads us to the optimal mindset of the participants in the review.  The developer of the code that is to be reviewed needs to have the mindset that the goal of a review process is to make the code product better.  Nobody likes having somebody point out a flaw in their work, but it is much better to find it during a code review than it is to have that flaw be exposed in a way that makes the evening news!  It is important to not take these problems personally.  In many cases, as a software developer, you are working on code that is owned by some other entity, such as the company you work for or some other organization.  It is useful when the code is being reviewed to have the mindset that the code, with all of its flaws, does not really belong to you but to the company you work for.  

Another aspect of the developer mindset is next elaborated. When working on the construction of code, or any other technical artifact such as a user story, it is well known that you reach a point when it is hard for you to improve the code - you may find yourself glaring at the code not understanding a thing.  This is a good time to ask others to look for problems in the code.  You get a fresh eye and a new perspective that will probably increase the chance of finding a problem. 

Yet another aspect of the developer mindset is being intentionally open and receptive to the review process.  That attitude facilitates the review process and ease the process of identifying issues.  The developer can confidentially do that as we separate in this style of reviews the identification of issues and their resolution.  In fact, issues that are recorded do not stand for a commitment or admission that something needs to be corrected or fixed.  Having that in mind ease the tension on the developer and enable her to relax and open up to the review process.

Next we turn to the reviewer to describe the ideal reviewer mindset.  As a reviewer, any problem that you find when reviewing a segment of code are not flaws in the person who wrote the code, they are flaws in the code itself.  You never want to insult or offend anybody when making comments.

%TBC - make the reference to code developer consistent throughout

One might ask where critical thinking fits into a code review.  Critical thinking can be defined in various ways.  The authors prefer the following simple definition: critical thinking is where one analyzes, questions, and interprets material and draws conclusions from that.  It is probably obvious that the review process forces one to analyze the piece of code that is being reviewed.  Asking questions about the behavior of that code is not only a technique for helping the reviewer to understand the code, but also to challenge any preconceived notions about the behavior of that code. It is also up to the reviewer to interpret the behavior of that code for accuracy.  All of this is done with the goal of drawing conclusions about the code being reviewed: Does it work properly or not?  Can it be done more efficiently or not?  If you have a good code review process, the idea of critical thinking enhances the review process.  The goal of a code review is to improve the code to make it the best that it can be, and if the code stands up to analysis and questions, then it hopefully will have high quality.  On the other hand, if the reviewers are argumentative or do a poor job of analyzing the code, or the code developer treats questions as a personal attack, the review will suffer.

An important aspect of the mindset of a reviewer is what their goal is.  It might seem like the goal is to prove that the code has no errors!  Although that is a nice idea, in practice this is impossible to prove.  If you knew that the code had an error, you would have fixed it!  And yet, bugs escape the development and test processes every day, only to be found by regular users.  You should assume that every piece of code has bugs.  The goal of a review should be to find as many bugs as possible, and fix them so that they are not found by your users.  

%TBC
%\begin{itemize}
%    \item General rules - budget time, prepare, evaluate after review
%    \item Mindset - find bugs, not prove absence of errors
%    \item Talk about hats?
%    \item Roles in a review
%    \item Need to gauge effectiveness
%    \item Code developer (owner) mindset
%    \item Stick to technical issues, not style.  Log issues, don't resolve them.
%    \item Reviews as education
%\end{itemize}

\chapter{Review preparation}

There are several considerations for holding an effective review and as a consequence planning the review is an important first step, especially for a project that can only afford to devote limited time and resources to a review.  

The first consideration is what should be reviewed?  This might seem like it would be simple to determine what code is to be reviewed, but it is often not simple.  Resist the temptation to bundle small, unrelated pieces of code together in a single review for convenience.  Reviewers tend to lose focus when switching between different pieces of unrelated code.  Also, comments raised on a prior piece of code may serve as a distraction, because a reviewer, or even the code developer, is still thinking about an issue from the prior piece of code that was reviewed.  

In addition, choosing the part of the code to review should be driven by what we are concerned about.   For example, the potential code to review has a server side function and a client side function and we are fundamentally concerned about the correct functionality of accessing the database on the server as the database engine was recently changed, then reviewing the part of the code that accesses the new database engine gets higher priority and might be the focus of the review activities. 

A review that is too large in scope can also be a problem.  No matter how efficient your review process is, it does take some time to review the material. It is better to break up a large amount of code into several smaller reviews.  This helps to keep the reviewers focused on the task at hand and from becoming bored.  This also helps to keep the reviewers from becoming overwhelmed with the amount of material to review.

Some statistics indicate that people will review code at a certain more or less constant pace.  As a result it may be useful to gather statistics on your review team and attempt to determine the pace in which effective reviews are achieved in your setting.  Having established a base line one can think quantitatively on prioritizing the potential material for the review.  For example, if we know that 50 lines of code take an hour to review in our setting and the material includes 250 lines of code, we require 5 hours to do it well.  If we only allocated an hour for the review of the material it is probably better to prioritize according to our concern and focus on part of the code that we are more concerned about and can be reviewed with high quality within an hour. 

%TBC - should we talk more about concerns here or should we point to a later chapter that elaborates on handling time constrants.

Another consideration is who should attend the review. Obviously, the code developer is required.  Any person who is an expert or has significant experience in the area of the code being written or updated should be considered as well, but you should not exclude somebody just based on a lack of expertise or experience.  Reviews can be a great source of education for newer members of a development team.  Also, in some cases it may make sense to include outside stakeholders.  For example, when creating a new internal service interface that will be exploited by a separate development team, it would make sense to include somebody from that other team.  Likewise, when that other team starts exploiting the service, it would make sense to invite the developer of that service to their review.

There can be too many people invited to a review.  At some point, there are diminishing returns simply because so many people want to be involved in the discussion.  Individual teams will find the best number of invitees, based on experience performing reviews and understanding the personalities and skill of the various reviewers, but a good initial guideline is no more than 5 to 7 reviewers.

Nobody should be invited to a review just to provide prestige or just for their own curiosity.  Anybody who is invited should provide value by helping to improve the code through their review or improving their understanding of the system.  

Yet another consideration is what materials to include in the review in addition to the code being reviewed.  Obviously, the code to review is necessary.  However, design documents, external documentation, documentation of external standards that the code must conform to, compiled output, and test output or traces are all examples of information that might be useful to include in the material being used in the review.  In general, if it can help the reviewers understand the code and/or find problems, it is probably worth including.

\begin{exercise}
\label{compare}
The following statements are capturing how different material can help in finding problems in the code.  Choose the statements that are correct.
\begin{enumerate}
\item Standard can be used to find problems in the code as we can compare the expected behaviour defined in the standard to the behaviour implemented in the code.
\item Test definition can help review the code as they may indicate missing tests that will find problems in the code.
\item Long description of code standards are highly helpful in finding problems in the code.  Reading hundreds of pages of code standard and comparing it to the code is a very effective way of finding problems in the code and should be always followed as first priority in code review.
\item Other implementation of the design patterns that are implemented in the code may help find problems as they may be compared to the implementation of the design pattern being reviewed.
\end{enumerate}
Solution: \ref{compare_solution}
\end{exercise}

%TBC - an expert reviewer - someone who is an expert in aspect of the code we are concerned about. 

Up to now we have discussed what part of the code to review and how to select the review participants. 
One last consideration in the planning phase is the review technique, or techniques, to use in the review.  We will cover several review techniques later in this book.  However, if you can identify concerns you might have with the code to be reviewed in advance, it might make sense to use one or more specific review techniques in order to focus the review and get the most value.  Some of the review techniques are more suitable for certain concerns which we will deep dive into when we discuss the review techniques.  To give an advance example in order to make the idea concrete, if we are concerned about interfaces and their correctness we will focus the part of the code that defines the interfaces but in addition we may choose to apply a review technique called "contract reviews" which is more suitable for interface checking and will be described later in this book.  Some of the techniques we will describe involve advance preparation as well as a review meeting.

%TBC
%\begin{itemize}
%    \item Planning: Attendees, artifacts, additional materials, review techniques
%    \item How much testing prior to a review?
%    \item Identify diminishing returns in a review - when to stop reviewing and ship it!
%    \item Need to talk about the group discussion aspects vs. independent review
%\end{itemize}

\chapter{Review techniques}
\section{Fagan reviews}

Fagan inspections, or Fagan reviews \cite{fagan}, are generally considered to be the first formulated technical review process.  This inspection technique was first formalized in the 1970's.  The technique has evolved since it was originally introduced. In spite of that we will provide a general overview of the technique since most of the other review techniques that we will cover in this book use elements of the Fagan review methodology.  We do not attempt to give a precise description of the original Fagan review methodology here but instead describe a version practiced by the authors. 

%TBC reference to the techniqe described below as clasical techncal reviews or in short technical reviews 

At the time the original Fagan review methodology was introduced, the waterfall development model was the most popular software development model.
The waterfall model has several development phases, and each phase must be fully completed and approved before moving on to the next phase.  

Although the phases for a waterfall development project can vary from project to project, a common example is as follows:

\begin{figure}
    \begin{tikzpicture} [every node/.style={rectangle,draw=black,align=center}] 
        \node(planning){planning}; 
        \node(design)[right= of planning]{design};
        \node(implementation)[right= of design]{implementation};
        \node(testing)[right= of implementation]{testing};
        \node(maintenance)[right= of testing]{maintenance};
        \draw[->] (planning.east) -- (design.west);
        \draw[->] (design.east) -- (implementation.west);
        \draw[->] (implementation.east) -- (testing.west);
        \draw[->] (testing.east) -- (maintenance.west);
    \end{tikzpicture}
    \centering
    \caption{Waterfall development workflow}
\end{figure}
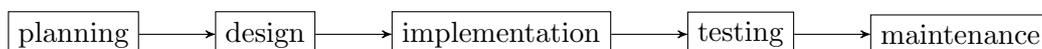

\begin{itemize}
    \item A planning phase gathers the requirements and objectives of the project.
    \item The design phase creates a formal document, called a design document.  This document describes the how the project will implement the necessary requirements or objectives of the project, including descriptions of the application's behavior.  It may also describe potential design trade-offs or other decisions that will affect the behavior of the application.
    \item Implementation is the phase of actually building the software to match the design specification.
    \item A testing phase would verify that the application actually implements the description in the design specification and meets the requirements for the project.
    \item Finally, the maintenance phase extends for the life of the software and includes diagnosing and correcting problems in the software.    
\end{itemize}

Note that the above process is characterized by capturing "what" the system needs to do, namely the requirement definition stage and then how this is going to be realised which spans over the design and implementation stages.   Interestingly the classical technical review process quickly goes over similar stages to recap what the code is supposed to do, understanding its design and then conduct the code review. The process of recapping the what and the how as preparation for the review process is agile and fast but also different in the sense that it focuses on which parts of the code are more likely to have problems.  In details following are the stages of the technical review:
\begin{itemize}
    \item A planning phase involves determining what code artifact should be reviewed and who should be involved in the review.  This phase also involves assembling the materials needed for the review and scheduling the review.
    \item %reread: updates for "can we some how emphasize that we also learn during the overview of concerns that the review needs to focus on", also moved some info down to a later paragraph
    An overview of the artifact to be reviewed is done.  We will describe the overview in more detail later in this chapter, but this step is a recap of the requirements and design of the component being reviewed.  
    
    Also, the various individuals participating in the review, including the developer of the code to be reviewed, may raise concerns or questions about the code artifact to be reviewed.  These concerns may be used to direct the reviewers to specific focus areas for the review.
    
    Technical reviews also assign roles to the individuals participating in the review, and these roles may be assigned to the participating individuals at this time.  We will discuss these roles shortly.
    
    \item The preparation phase gives the reviewers the chance to review the artifact and perform their own investigation, noting questions or concerns.  This is typically done independently by each participant.
    \item The inspection meeting brings all of the participants together to review the artifact as a group, with questions or concerns being documented.
    \item In the rework phase, the code developer then follows up on all of the questions or concerns, potentially making changes to the code artifact.
    \item Finally, in the follow-up phase, it is necessary to ensure that all of the questions and concerns have been resolved to the satisfaction of all of the participants in the review.
\end{itemize}

%reread next two paragraphs for comments:
%do I want to also mention overview can cause restart too?
%Mention that this process is applicalbe to any tehnical aritifact not only code. For example, use cases.
At any phase, it may be determined that the phase can not be completed satisfactorily.  In that case, it may be necessary to restart the review process from the planning phase.  This is most common in the rework phase, when it is often apparent that the questions or concerns that were raised necessitate reviewing the artifact again.  This can also happen during the overview phase, since concerns may be raised that cause the code developer to decide that additional changes are needed before performing the review, or even that the code is not ready to review.

In this book, we are focused on performing technical reviews for code artifacts, but technical reviews are not limited to reviewing code.  This review flow can be easily adapted to reviewing design documents, use cases, test plans, or documentation.

%TBC - maybe create a digram of requirement design implemntation and the corresponding technical review stages.

Above we mentioned that the technical review process recaps the code to be reviewed, matching the waterfall model by going over the requirements and design in the overview phase and the implementation and code in the preparation and review meeting.  You can begin to see another synergy between the technical review process and the waterfall development model at this point.  In both cases, each phase of the process has a well-defined set of inputs and outputs, and it is necessary to complete each phase before moving on to the next phase.

%reread for "elaborate on what do we mean by an overview.  Conccretely it includes what the components is doing, the main parts of the components and how they are combined to achieve the use cases that the component supports."  Also moved some text from above.
The overview should include a number of items.  It should describe the overall function that the component performs.  It also should include the main pieces of the component and its overall logic, as well as interactions with other components or functions that may be relevant to the review.  It also should describe relevant use cases and how the parts of the component fit together to support those use cases.  This is, in fact, a recap of the requirements and design of the component to be reviewed.  The purpose of the overview is to provide context for the code artifact to be reviewed, since in many cases, the review will only include a part of the component rather than the entire component.

As previously discussed, the participants in the review may perform one or more roles in the review process:
\begin{itemize}
    \item The author, or owner, is typically the developer who created the code to be reviewed.  This person is responsible for resolving any issues discovered during the review process.
    \item The moderator is in charge of managing the inspection process from start to finish.  The moderator is in charge of planning the review, choosing participants and assigning roles to them, and keeping the review process on track.
    \item The scribe records any issues found in the review meeting.  They are responsible for making the documented issues available to the review team as soon as possible after the inspection meeting, typically within a day of the review.
    \item The reader will paraphrase the code for the reviewers.  We will go into paraphrasing in more depth in a following section, but this basically means explaining in plain terms what the individual lines of code are doing.
    \item The reviewers, or inspectors, will review the artifact to find errors, omissions, inconsistencies, or other issues.
\end{itemize}

Note that it is common for participants to perform multiple roles in a review.  For example, the moderator often performs the role of the scribe, and all of the participants will typically act as reviewers.

One important aspect of technical reviews is the need to capture some data to determine if the review is effective.  One piece of data to capture is the amount of time spent on the review process, not just including the time spent as a group in the review meeting, but also the time spent on individual tasks such as preparation.  It is also important to capture the number of issues, their severity, and the types of issues.  This data can be used to determine the overall quality of a particular component or application.  Finally, the number of lines of code being created or changed, as well as the number of lines of code being updated or written as part of the rework phase, can be used to give an indication of whether or not it is necessary to hold a re-review of the code artifact.

%reread next two paragraphs for: "1. our experience shows that technical reviews for an organization that are consisitenlty conducted will have a tpycal rate.  2.  Either point to a reference of a typical number or give a typical number that we feel comfortable with in our pratice 3. You shold collect your own data and determine a number that is correct ofr you pratice." and minor other rework.
In the experience of the authors, we have found that classical technical reviews for an organization tend to have a consistent inspection rate.  In our organizations, we have found that classical technical reviews typically have an inspection rate of approximately 100-125 lines of code per hour.  However, this rate will vary from one organization to another, based on the experience of the reviewers, the complexity of the component, and the maturity of the code.  This is another piece of data to capture as part of the review process and can be used as part of the review preparation process for future reviews to determine how much time is needed for a particular review.

Regardless of the actual inspection rate you find in your organization, for large software projects of many thousands of lines of code, if not millions of lines of code, a classical technical review with an inspection rate measured in hundreds of lines of code per hour will quickly become impractical.  In a classical waterfall development cycle, this could require hundreds of hours to review an entire software project.  When applying a classical technical review in an agile setting, each iteration will typically require some amount of review, which can have an additive effect on the time needed to review as later iterations will often require re-review of code delivered in earlier iterations.

However, there are many aspects of classical technical reviews that can (and should) be used as part of a good review process regardless of the development model that is used.  As an example, having review comments documented for later follow-up is valuable for several reasons.  First, it ensures that the review comments are actually addressed and gives the reviewers the chance to agree or disagree with the way in which these comments were addressed.  Second, these documented comments and their responses can be used in the future when questions inevitably arise about why a particular piece of code was implemented the way it was.  These comments often provide some insight into low-level implementation decisions that were made during the development cycle.  Finally, holding a review meeting and being able to prove that comments were addressed is often a documented requirement of the development cycle and may be referred to during later audits of the development process.

%TBC - should we/do we have a separate section on review planning?  This might be already covered in the Review Prep chapter, we might want to re-read to see if it's complete.

\section{Paraphrasing}
The Paraphrasing review technique is one of the simplest review techniques to understand.  It is also the most common technique in use because it is a general purpose technique that can be used in most situations.  To put it simply, it is the act of "paraphrasing" or explaining in common language what a particular piece of code is doing.  When a review is conducted the reviewers should understand the code and as a consequence in most situations paraphrasing can be applied. For the reviewer that do not understand the code it quickly brings them up to speed making the review process more effective. 

As an example consider the following piece of code:
\begin{lstlisting}[language=C]
if (x > 0) {
    print(status(a[x]));
}
\end{lstlisting}
The simplest way to read that is "if x is greater than 0, then print whatever value is in the status array for index value x."  However, that does not really explain what the code is doing.  A better way to read that is "x contains the state of the device and any non-zero value indicates an online state, so if x is greater than 0, we should print the state of the device using the string in the status array."  The second option is what we refer to as paraphrasing.

In an ideal world, the code to be reviewed would be self-contained and readable in a way that it would stand by itself and not require much rewording to explain using common language.  In the real world, code is often obfuscated, variables do not have descriptive names, and comments are insufficient, so some amount of rewording or explanation of the code is likely to be needed.   

The advantage of paraphrasing is that it helps facilitate understanding.  If a particular piece of code is hard to explain using clear language, it can be an indication that it is not well implemented and as the team maintains it they are more likely to make mistakes as it is not readable.

\begin{exercise}
\label{paraphrasing}
    Consider the following code snippet:
    \begin{verbatim}
        if ((numberOfGroups > 0) and (answerWasGiven() == true)) then 
          doSomething()
    \end{verbatim}
    Which of the following is the best way to paraphrase this code snippet:
    \begin{enumerate}
        \item “We check that numberOfGroups is bigger than zero and that answerWasGiven() returned true, and if so call doSomething()”
        \item “We check that we were able to find at least one group and that an answer was given, and if so we do something.”
    \end{enumerate}
Solution: \ref{paraphrasing_solution}
\end{exercise}

\section{Obligations and contracts}
In the previous section, we described paraphrasing, which is a "general purpose" review technique that can be used in any situation.  We now begin to discuss several techniques that are specific to certain situations.   Such more specific review techniques are aimed at addressing a concrete concern and are geared towards finding problems that are a realization of the specific concern.  For example one might be concerned that the code being reviewed has race condition or deadlock.  Alternatively, one may be concerned that the interfaces of the code are not well defined or agreed upon by the different stakeholders that are using them.  

We will begin our discussion of specific review techniques with the obligation review or contract review that indeed focused on interfaces and their correctness. 

First, let us describe what we mean by an obligation.  Typically, an interface consists of two pieces: the caller of a function and the function itself.  An obligation defines what must happen, or must not happen, at some point before or after the caller invokes the function.  This can also be thought of as a contract between the caller and the function.

For the purposes of considering obligations, the function can be an internal function within the product or application, or an external API that can be used by any caller.  For internal functions, both the caller and the function owner are probably available for the review, but for an external API, the caller(s) may not be available to take part in the review process and some effective representative should be identified. 

Obligations for the caller of a function can fall into two categories.  The first is a pre-condition, which defines what the caller should do prior to calling the function.  One example of a pre-condition would be that the caller should open a file successfully before calling the function to read that file.  The other category is a post-condition, which defines what the caller should do after calling a function.  An example of a post-condition is that if the function that reads a file fails with a specific error condition then the caller should invoke a recovery routine to repair the file.  The recovery routine, among other things, will typically free resources that were allocated for the operation that failed.

Another category of obligations has to do with the actual semantic of the function being called.  For example, the function may guarantee that the number of references to a file by processes that are manipulated the file is always the same as the number of file meta blocks that have the incore representation of the file.  That equality is sometimes refers to as an invariant.  The semantic of the function is to always keep that equality.  Another more commercial example of an invariant is how much money a person has in his bank account.  This amount has to accurately match the transactions that have been executing against the account.  The invariant here may be stated figuratively as "money does not grow on the trees".  Checking such invariants applies to a set of interfaces and not only one interface.  In the account example, all operations that withdraw and add money to the account should be reviewed together to assure that the invariant is kept.   

The intent of an obligation review is to review the code before or after calling the interface to ensure that both the caller and the function meet their obligations as well as that the invariants are kept.  To continue with the example of reading a file, an obligation review might verify that the caller of the read function closes the file after reading not just in the successful path, but any error paths as well.

\begin{exercise}
\label{obligation_signals}
Signals are a mechanism that enables the change of the control flow of a process in Unix (see \href{https://man7.org/linux/man-pages/man7/signal.7.html}{link} for details).  Once a  signal is raised against a process, the execution of the process may be stopped and may be resumed in a new function called a signal handler. When we say that we need to check obligations along all of the error paths we mean that it includes:
\begin{itemize}
    \item The calling of the signal handler while the function is executing.
    \item The calling of the signal handler under different types of signals while the function is executing.
    \item None of the above, handling of signal is out of scope for obligation reviews. 
\end{itemize}
Solution: \ref{obligation_signals_solution}
\end{exercise}

This review technique is valuable when introducing a new interface, modifying an existing interface, or creating a new use of an existing interface.  Interface defects often account for a small number of defects in a project, but they tend to be hard to find until late testing phases and can be expensive to fix.  This review technique can also make sense when changing error recovery handling, to ensure that resources have been properly cleaned up.

The approach to an obligation review is to consider all of the obligations of a function and its caller, and review to make sure that the obligation is handled in all paths.  The first step is to ensure that all of the obligations are documented in an appropriate location, because this information is only useful if it can be found.  If possible, the obligation should be documented in the code in order to make it obvious to developers who are modifying that code.  In the case of an external function that can be used by many different callers, the obligations should be described with the documentation of that function.  

The amount of documentation needed for an obligation will vary, but it should be short enough that it can be easily read and yet long enough to contain any necessary information.  Some obligations may be not be explicitly specified on the individual function.  As an example, if a package is not thread-safe, it would make sense to document once that the package is not thread-safe rather than documenting it on every function that the package supports.  On the other hand, if the overall package is thread-safe but specific functions are not thread-safe, then each function should state whether it is thread-safe or not.

If possible, obligations should be enforced by the code, but this does not take the place of having good documentation.  The caller of a function should not have to find out that an obligation exists through trial and error.  Compile time checks for obligations are the best choice since they do not impact code execution and do not require run time testing to verify that the obligation is met.  However, some obligations can only be verified at run time, but these require testing for verification that the obligation is met in all cases.

Once all of the obligations of a function are documented, the callers of the function should be reviewed to ensure that they meet the obligations on all paths.  Many times only the mainline path of the function is considered but error paths are not.  

\begin{exercise}
\label{obligation}
    We are given a resource R and its associated global use count i.   There are three threads using the resource R.  Initially the use count i is set to 0.  Each thread calls obtain(R) to use the resource.   obtain(R) atomically increments i by one.   The thread is then required to release the resource by eventually atomically decrementing the use count by one by calling release(R) (an obligation).  

    Threads 1 and 2 have the following code:
    \begin{verbatim}
        obtain(R);
        ...
        release(R);
    \end{verbatim}
    Thread 3 has the following code:
    \begin{verbatim}
        obtain(R); 
        ...
        if (i > 1) then
            release(R);
    \end{verbatim}
   
   Which of the following interleavings break the obligation?
   \begin{enumerate}
       \item 
       \begin{verbatim}
t1.obtain(R);
t2.obtain(R); 
t3.obtain(R); 
if (i > 1) then t3.release(R); 
t1.release(R); 
t2.release(R);
       \end{verbatim}
       \item 
       \begin{verbatim}
t1.obtain(R); 
t2.obtain(R); 
t3.obtain(R); 
if (i > 1) then t3.release(R); 
t2.release(R); 
t1.release(R); 
       \end{verbatim}
       \item 
       \begin{verbatim}
t3.obtain(R); 
if (i > 1) then t3.release(R); 
t2.obtain(R); 
t2.release(R); 
t1.obtain(R); 
t1.release(R); 
       \end{verbatim}
   \end{enumerate}
Solution: \ref{obligation_solution}
\end{exercise}

Sometimes it it hard to determine the different possible code paths in which an obligation should be checked.   To see this consider the following exercise.  

\begin{exercise}
\label{obligation_f}
A function \verb|f(i)| accepts an integer \verb|i| and increments it by \verb|1| only if some external server is up and running.  The code implementing the function may look like 
\begin{verbatim}
int f(int i) {
  if(serverIsRunning() == True) 
    return(i+1); 
  else 
    return(i);
}
\end{verbatim}
The function \verb|serverIsRunning()| may take an exception.  In such a case, an exception handler is called to complete the operation.  Which of the following alternatives is correct:
\begin{enumerate}
    \item It is enough to check that \verb|i| is incremented when the function is returned in order to check its obligation.
    \item One needs to check that \verb|i| is not incremented when \verb|serverIsRunning()| takes an exception.
    \item The obligation is of \verb|f()| is not well defined as it is not clear if the server is up and running or not if \verb|serverIsRunning()| takes an exception.  The correction of the obligation that takes this case into account should be checked both for the case that the function returns normally and the case that the function takes an exception.   
\end{enumerate}
Solution: \ref{obligation_f_solution}
\end{exercise}

\section{Bug patterns and checklists}

Checklists are a very simple and time-honored technique used in many different industries to help reduce defects or errors.  A checklist is just a list of items to keep track of or remember.  As an example, rather than trying to remember all of the items that you need to purchase at the grocery store, you might make a list of those items so that when you are at the store, you can make sure that you have all of those items in your cart before you leave.  

In the context of software engineering and code reviews, a checklist is a list of bug patterns to look for when reviewing code. A bug pattern is a documented error that commonly occurs, including symptoms of the error and potential ways to fix the error.

Creation of checklists can be an excellent technique for capturing known bug patterns if done well.  However, a poorly implemented checklist can be difficult to use, and that often results in a checklist that is ignored.

A checklist should follow a common template for each bug pattern.  The template should be adjusted to the needs of the development organization, but the authors recommend the following starting point for the template:
\begin{itemize}
    \item Short pattern name
    \item Longer description or root cause
    \item List of symptoms
    \item List of ways to prevent and/or fix the issue
    \item Review activities, including both what to look for in a review
    \item Test approaches: how to test for the bug pattern
    \item Optional coded example of the bug pattern
\end{itemize}

Generally, a checklist should be of a manageable size so that the reviewers can review the items on the checklist prior to the review and remember them during the review process.  A good starting size is 10 items.  In contrast, a document with 500 code pitfalls that is meant to be used as a check list is clearly useless.  The document could be useful as a reference and be utilized for learning purposes but once you are attempting to apply it as a checklist, you need to reduce it to a small number of bug patterns you should make sure are not present in the code being reviewed and are known to be important to focus on. 

A good checklist should be customized to a meaningful scope.  The scope should be no larger than a component.  A scope such as a company-wide checklist that encompasses many components or products will tend to have limited value since it is too generic.  The wider the scope of a checklist, the more likely it is that it will need to contain more items, which makes it harder to remember all of the items on the checklist during the review.

An appropriate scope might be as small as a checklist tailored to a single source file or set of source files, or as large as a checklist covering an entire component.  In many cases, it might make sense to have multiple checklists, such as a general checklist for the component, as well as checklists covering specific sub-components.

%TBC - Scott: should we talk about pitfalls in non-abstract (see Eitan's decks):
%TBC - possible reference https://www.researchgate.net/publication/220950728_Concurrent_Bug_Patterns_and_How_to_Test_Them

%TBC Scott: consider size, scope, technical attributes of the code (serialization, etc.)

It is important to consider the technical attributes of the code when developing a checklist.  This can include considerations such as concurrency, where multiple processes are running in parallel, and synchronization, where those parallel processes need to use a synchronization method such as a lock or semaphore to prevent the parallel processes from updating a common area at the same time.  This can also include error recovery and exception handling for errors that occur during processing.

We will consider synchronization as an example of technical attributes to be considered in a checklist. 
A simple component that does not require synchronization would have no reason to cover synchronization in a checklist.  However, if a component uses a single common lock for synchronization, then it is enough for the checklist to include a bug pattern that describes the conditions for which synchronization is needed.  It might also take into account the synchronization primitives and whether they have any specific requirements, such as considering if error handling needs to explicitly release the synchronization if an error occurs as opposed to the system releasing the synchronization automatically on an error.  Instead, if the component uses multiple locks, each for a different purpose, then the checklist might need to consider different bug patterns, covering the various locks and what resources they synchronize as well as any hierarchy between the different locks (or order in which multiple locks should be obtained).  These patterns would identify incorrect use of the locks, such as using the wrong lock to synchronize a piece of data which would effectively leave it unsynchronized, or potential deadlock situations where a process obtains a lock and then has to wait for a second lock which is held by another process that is waiting for the lock held by the first process.

\begin{exercise}
\label{pitfalls}
Programming pitfalls can be viewed in concrete or more abstract way.   In fact, an abstract principle of how people make mistakes may be mapped in this work to a generic review process that is then actually captures and attempts to reveal a range of more concrete programming pitfalls.  To give a concrete example contract review is interest in resource handling and whether they are correctly released along error paths. This may map to several concrete programming pitfalls that realise the generic resource releasing pitfalls in different ways; all of which can be found using the contract review but also can be found by directly inspecting and search for the specific programming pitfall.  In the following choose which programming pitfalls map to the generic correct releasing of resource along error paths concern.
\begin{enumerate}
    \item An out of boundary array access.
    \item Obtaining a lock and then calling a function that takes an error path the does not release the look.
    \item Obtaining a lock and then calling a function that takes an error path the does not release the look.  Obtaining a pointer to an object and then not releasing the pointer along an error path.
    \item Error paths always lead to the termination of the process or thread so in modern languages releasing of errors along error paths is not a concern.  
\end{enumerate}
Solution: \ref{pitfalls_solution}
\end{exercise}

In some cases, it also might make sense to have a checklist tailored to a specific review.  
For example, if you are adding a new data structure to a component, including a checklist element that reminds you that every time that fields in the structure are referenced, you need to first verify that the structure exists by checking that the pointer to the structure is non-zero.

Checklists should be focused on types of errors rather than style considerations.  It can be useful to have a style guide for the component or organization, but a technical code review should be focused on finding defects rather than correcting style issues.  In addition, given the current state of the technology many style guidelines can be easily enforced using some automatic processing of the code which underscore that it is best to focus on finding technical problems in the code instead. 

Checklists should also be reviewed periodically.  Unfortunately, it is common to add new patterns to the checklist, but never remove old patterns, resulting in a checklist growing to an unusable size.  Once a checklist grows too large, old patterns should be reviewed to determine if they are still relevant.  It might be appropriate to break up a checklist into separate checklists with a smaller scope.  For example, some patterns in a large component-level checklist might be moved into different sub-component-level checklists that are of a more manageable size.  Another way to organize check lists, in addition to organizing it by component structure, is to organize the check list by type of programming pitfalls such as concurrency issues, memory leaks, interface issues, etc. 

\begin{exercise}
    \label{checklist}
    Checklists represent programming pitfalls.  When setting up a review, you need to choose a check list for that review so that:
    \begin{enumerate}
        \item The checklist will be complete and represent all known pitfalls.
        \item The checklist is focused on the risks that apply to this review and is short.
        \item The checklist focuses on enforcing important standards.
    \end{enumerate}
Solution: \ref{checklist_solution}
\end{exercise}

There are many good examples of development and code review checklists available.  One very well known checklist is the OWASP Top Ten which can be found at\\
\hyperlink{https://owasp.org/www-project-top-ten/}{https://owasp.org/www-project-top-ten/}.

In the rest of this chapter, we will be covering some common programming pitfalls in detail and discuss how they are typically found in reviews and testing, and root cause and appropriate fixes for those pitfalls are made.  These details will serve to enhance your understanding of programming pitfalls and serve as a baseline for creation and customization of pitfalls and checklists for the reader's context.  Whenever possible, we will also highlight the level in which the pitfall is occurring, as some programming pitfalls are really design issues rather than coding issues, and some are actually coding pitfalls.  The following section is more technical and can be skipped at first reading if the reader would like to obtain an overview of the review techniques presented in this book first.

\subsection{Concurrency pitfalls}

Concurrency programming is hard.  It requires the understanding of interacting processes over time to achieve a common goal.  As processes interact they access shared resources or pass messages between them.  They may hinder the execution of each other as they compete on shared resources or wait in a cycle for something to happen.   In this section we detail major types of pitfalls related to concurrent programming.  

\subsubsection{Protect access to shared resources [\textit{protect-general}] }

When writing concurrent programs one should be aware of the need to protect the access to shared resources using the synchronization primitives provided by the environment.
Typical synchronization primitives include locks that prevent more than one process accessing a resource at the same time, 
volatile variables that ensure new values of the variables are visible to all threads, compare and swap operation primitives that make sure that when a value of a variable is checked and then changed nothing happens in between, and others. 

Clear and precise understanding of the synchronization primitives is essential to confident and correct concurrent programming.  For example, one should clearly understand what operations are atomic and what operations are not atomic. 
In Java, int operations are guaranteed to be atomic but long are not.  Unawareness of such details may lead to programming errors.  Another example is visibility of changes made by a thread to all other running threads. 

\paragraph{Programmer:} 

The programmer needs to identify shared resources and protect access to them using appropriate synchronization primitives.  Doing this is sometimes tricky as the shared resources may be implicit.  For example, the Linux \href{https://man7.org/linux/man-pages/man3/errno.3.html}{errno} used in system calls is process scoped and is a shared resource that threads will race on.  In addition the programmer should pay attention to strike a trade-off between safety and performance.  For example if the scope of a lock is bigger than it needs to be, processes waiting on that lock will wait uneasily for the lock to be released.  Thus, to improve performance the lock should only protect the actually shared resource access.  

Consider the following example in which we assume the only shared resource is the queue.  In the example the lock is scoped to include the preprocessing and the message printing.
\begin{verbatim}
lock
doing some preprocessing to make a decision if I'm accessing the queue
printing messages
access(queue)
unlock 
\end{verbatim}

Other processes may wait unnecessarily on the lock while this process performs the preprocessing and message printing, even though that processing does not require the lock.    

Similarly, if a single synchronization mechanism is used to protect several distinct resources, then this can cause unnecessary contention. Consider this example in which we have two threads using a single lock to access two different queues:

\begin{multicols}{2}
Thread 1:
\begin{verbatim}
obtain_lock(lock_A)
access(queue_1)
unlock(lock_A)
\end{verbatim}
\columnbreak
Thread 2:
\begin{verbatim}
obtain_lock(lock_A)
access(queue_2)
unlock(lock_A)
\end{verbatim}
\end{multicols}

Since only one lock is used to synchronize access to both queues, thread 2 is not able to obtain the lock to access \verb|queue_2| while thread 1 holds the lock, even though thread 1 is not accessing \verb|queue_2|.  Instead, using two separate locks, one for \verb|queue_1| and one for \verb|queue_2|, will allow access to each queue independently.

\paragraph{Tester:}
\label{protect-general-tester}
The tester needs to ensure that their tests are creating scenarios that result in contention on these shared resources.  For example, one thread should hold the lock while other threads are waiting for the lock to become available.  To do this, it is necessary to understand which methods or code segments can access the shared resources concurrently.

One possible approach is to run a sequential test in parallel.   This is sometimes referred to as multiplication.  To implement the approach you create several copies of the same test and run them in parallel each running in a different thread.   For example, if you test is writing to a file $n$ bytes you could multiply the test by running $k$ tests in parallel each one of them writing $n$ bytes to the file. We make the example concrete in the following pseudo code segment. 

\begin{lstlisting}[language=C]
writeToFile(int n, String fileName){
   buf = randomBuffer(n);  // obtain a random buffer of n bytes
   fd = open("w", fileName);
   if(fd == success){
        result = write(fd, buf);
        assert(result == success); //write is expected to succeed 
   }
   close(fd);
}
\end{lstlisting}

To multiply the sequential test above we create $k$ threads that execute the same write operation in parallel see pseudo code snippet below. 

\begin{lstlisting}[language=C]
fork k threads and have each thread execute:
    writeToFile(100, "myFile");
\end{lstlisting}

The multiplication approach can be enhanced in various ways.  One enhancement that immediately comes to mind it the size of the file being written.  You could have each thread randomly pick the size of the file from a  given range to make the test stronger.  In addition, one may want to multiply sequential test of different type that access the same shared resources.  For example, another sequential test might be reading the same file (see pseudo code below).

\begin{lstlisting}[language=C]
readFile(int n, String fileName){
   buf = randomBuffer(n);  // obtain a random buffer of n bytes
   fd = open("r", fileName);
   if (fd == success) {
        result = read(fd, buf);
        assert(result == success); //write is expected to succeed 
   }
   close(fd);
}
\end{lstlisting}

Now we may want to execute the reads and the write in parallel as the following pseudo code suggest. 

\begin{lstlisting}[language=C]
fork k threads and have each thread execute:
  if(randomchoice == write) { // randomly choose between 
                              // reading and writing
      writeToFile(100, "myFile")
  } else {
      readFile(100, "myFile")
  }
\end{lstlisting}

One other point about contention is that contention may be explicit and requires grey box understanding of the implementation.  For example, the file open process accesses a table of system files when it opens a file so even if file operations are executed on different files they still contend on a shared resource.  Knowing that it is interesting to also multiple read and write tests that access different files.  If we create a set of files $FILE = \{file_1, ..., file_l\}$ then we end up with the following multiplication.  

\begin{lstlisting}[language=C]
fork k threads and have each thread execute:
  fileName = random(FILE) // choose a file from the set of 
                          // files at random
  if(randomchoice == write) { // randomly choose between 
                              // reading and writing
      writeToFile(100, fileName)
  } else {
      readFile(100, fileName)
  }
\end{lstlisting}

Indeed multiplication can create complicated and interesting scenarios only limited by your ability to understand the semantic of the system under test and define you expected results to the multiplication scenario.  This should be many times determined through discussion with the programmer and typically reveals scenarios that where ignore during the development and by just highlighting them problem are found even before the tests are executed. 

It is also important to understand the choice of synchronization primitives that are in use.  In reading the prior pseudo code examples of reading and writing files, we did not fully describe how our file synchronization protocols worked.  There was an implicit assumption that when opening a file in write mode, an exclusive lock on the file is obtained so that no other thread could write to the file, and closing the file would release that lock.  You might also assume that when opening a file to read it, a shared lock on the file is obtained that would allow other readers of the file to read it in parallel, but prevent more than one writer of the file from accessing the file as the writing between writers is exclusive.  If this is the case, then as a tester you would expect that multiple threads that are reading the same file at the same time should be able to execute in parallel.

However, the behavior is different if the open process requires an exclusive lock on the file in order to read it.  This might be the case if the files inherently require sequential access, such as reading a file from a tape device.  In this case, if you find during testing that two threads are able to perform file operations at the same time, that is a clear indication of a synchronization problem.

%TBC - Eitan, should we talk about ConTest here?

Effective communication between the programmer and tester is always important, but in the case of concurrency testing, it is essential.  The programmer often understands the synchronization primitives and how they can be used, and has insight as to the scenarios in which contention can occur.  This information can save the tester valuable time in developing test scenarios.  Likewise, as the tester investigates potential test scenarios, they may uncover shared resources that were not properly synchronized, or incorrect use of the synchronization primitives.  This can help the programmer correct these issues even before executing the tests.

%TBC - Eitan, we covered multiplication very well here, but should we consider an example where multiplication does not make sense?  For example, interaction between a client and a server.

%TBC - Maybe not part of this pitfall, but what about concurrency and performance/throughput?  Scott - this is the SYSZTIOT batching scenario we fixed for Db2.  Maybe too weird to discuss in this book.

\subsubsection{Avoid errors by understanding the synchronization primitives [\textit{precise}] }

Another issue with writing concurrent programs is the need to use the synchronization primitives correctly in all environments.  A lack of understanding can manifest itself in several different ways.

It is necessary to understand how the code that is executed within the scope of a lock handles an exception.
If the lock is automatically released by the system before the exception handler receives control, then the exception handler may need to obtain the lock again if it is necessary to perform a recovery action against the shared resource.  If the lock is not automatically released, then it may be necessary to explicitly release the lock.

It is also necessary to understand how the synchronization primitive handles requesting a lock that is already held by that thread.  Most lock services will behave in one of three ways.  

The synchronization primitive may simply grant the lock that is already held by that thread.  This may have implications for obtaining and releasing the lock.  For example, an exception handler may be implemented to always obtain the lock to process some shared resource, and release that lock before the exception handler gives control back to the mainline processing.  If the mainline processing obtains the lock, and then an exception occurs, the exception handler will request the lock again and it will be granted as the lock was already held.  If the lock is released by the exception handler and control is given back to the mainline processing at a point that expects the lock to still be held, then the shared resource is no longer protected.  This type of lock is sometimes referred to as a "recursive lock".

Another possibility is that the synchronization primitive may reject the request if the lock is already held.  This is sometimes referred to as a "non-recursive lock".  If the synchronization primitive throws an exception when this occurs, then an exception handler that obtains that lock needs to understand if the lock is already held in order so that it does not generate another exception.

Yet a third possibility is that the synchronization primitives do not protect against the lock already being held.  In this case, if the thread already holds the lock and the lock is requested again by the same thread, the thread may wait on itself to free the lock, which will not happen.  This effectively creates a thread that is deadlocked on itself.

The above pitfalls often occur due to a lack of understanding of what responsibilities the synchronization primitives have and what responsibilities the user of these synchronization primitives have.  Obligation reviews can be a useful technique in uncovering these pitfalls.  That review technique ensures that responsibilities such as when locks are released are well documented and well understood, and that all of the paths through the processing, including error and exception paths, are reviewed to ensure that the locks are released and held at the appropriate time.

Another pitfall associated with misunderstanding of synchronization primitives is that locks are not atomic operations.  Specifically locks do not make the access to shared data atomic.  Instead, locks are used to implement atomicity through controlling whether or not threads execute a critical section in which the shared data is accessed.  They are as good as they are adhered to.  For example if you have two threads accessing a shared resource one holding a lock before the resource is accessed and another that does not then no protection is given to the shared resources.  This is sometimes referred to as a "gentleman's agreement", where every process that accesses the shared data must agree to access it using the same synchronization method.  If any process does not follow the agreement, the shared data is effectively not synchronized or protected at all.

Yet another pitfall has to do with visibility of changes made to the data.  When several layers of memory are implicitly used, a write to a data item by one thread may or may not be visible to another thread depending on implicit semantic of the writing and of other synchronization primitives execution. 
For example, if we update a variable $x$ to have the value of $1$ this may be written to memory that is scoped to a given thread and flashed to memory that is scoped to the program only when we unlock a lock that was held while write to $x$.  If the second thread is making a control flow decision based on the value of $x$ that may cause a bug even if $x$ was updated by the first thread, as the updated value of $x$ is not visible to the second thread.  

Such visibility problems are not limited to the relation between the variables to be updated and locking.  For instance some programming languages may have integers of type volatile (e.g., Java) to mean that their update will be forced to the joined memory scoped to the process.  If the programmer is not aware of this visibility issue he may type the variable in such a way that desired visibility of changes across threads is not achieved thus creating visibility bugs. 

Another example has to do with waiting on an event to occur.  Some synchronization primitives such as $pthread\_cond\_wait()$ in C and $wait()$ on an object instance in Java provide a systematic way to wait on an event.  One should carefully understand the range of events that the interface support waiting on and exactly what happens when the even occur.  For example, event being supported may include interrupts, signals, etc, the point being each event has its own semantic of what exactly happens when it occurs.  In addition, there are general considerations of the type of things you need to take into account to clearly understand the behaviour when an event occur.  For example, waiting on events is typically done while holding a lock and finding out that some condition has not occurred yet. When the wait occurs the lock is implicitly unlocked to enable other threads to advance in the hope that the desired condition will occur.  Thus, typically you expect to obtain the lock again implicitly before returning from a wait operation when an event occur.  Careful study of the wait operation semantic should reveal if this is indeed the case for each of the possible events that may occur.  In addition, an indication that the process event occur may reside in some variable with additional information of the exact event that occurred.  Again careful study should be made to understand the life span of that indication.   

Another point to note is the process is in a wait state and an event occurs the process may implicitly obtain the lock in a way that is not atomic.  As a result, the event that the process waited on on may no longer be in effect when the process has obtained the lock.   For example, if we are waiting for a file size to get to certain size, we may encounter an event that the file got to this size but by the time we reobtain the lock some of the file data may have been deleted resulting in the file being less than the desired size.  As a result, once we obtain the lock we should double check that the size of the file is indeed as big as we were waiting for it to become. 

%TBC - our review methdolgy also checks design and requiremnts
%TBC - our approach to bug pattern is unique - it is abstract and not language specfic - language examples are typically given in the excercises and examples. 

There are also considerations related to unexpected external events, such as an interrupt or signal occurring while waiting. As an example, when an interrupt occurs during a $wait()$ in Java, an exception is raised, but the interrupted flag is turned off so that the caller is not able to use that flag to know that the thread was interrupted.

\begin{exercise}
\label{pthread_cond_wait}
Consider the $pthread\_cond\_wait()$ C function in UNIX, \href{https://pubs.opengroup.org/onlinepubs/7908799/xsh/pthread_cond_wait.html}{link}, which is used to block a process until a condition is satisfied.  This function is called while holding a mutex lock.  When called, the $pthread\_cond\_wait()$ function unlocks the mutex and the thread blocks on a condition variable.  When the thread is successfully unblocked (presumably because the condition variable has been updated) the mutex is reacquired prior to returning from the $pthread\_cond\_wait()$ function.  For example:
\begin{lstlisting}[language=C]
    // Define the mutex and condition variable
    pthread_mutex_t my_mutex;
    pthread_cond_t condvar;
    // Define the predicate condition variable and initialize it to zero
    int condition_case = 0;

    pthread_cond_init(&condvar,NULL); // Initialize the condition variable
    pthread_mutex_init(&my_mutex,NULL); // Initialize the mutex lock
    
    pthread_mutex_lock(&my_mutex); // Obtain the mutex lock
    while (!condition_case) { // While the predicate condition is still 0
        pthread_cond_wait(&condvar, &my_mutex); // Wait on the condition 
                                                // variable
    }
    pthread_mutex_unlock(&my_mutex); // Release the mutex lock
\end{lstlisting}

If the thread is cancelled while waiting, what is the state of the system when the first cancellation cleanup handler gets control?  Choose the best possible answer:
\begin{enumerate}
    \item The mutex will be reacquired prior to the cancellation cleanup handler getting control.  The state of the condition variable is unknown.  The cleanup handler can safely proceed as if the mutex has been reacquired.  
    \item The mutex will not be reacquired prior to the cancellation cleanup handler receiving control.  The state of the condition variable is unknown.  The cleanup handler can safely re-obtain the mutex if needed.
    \item It depends on the cancelability state of the thread.  If it was set to \\   
    $PTHREAD\_CANCEL\_DEFERRED$, the mutex will be re-obtained prior to the cleanup handler receiving control.  If not, then the state of the mutex is not known and the cleanup handler must handle both cases, where the mutex is held and the mutex is not held.  In either case, the state of the condition variable is not known.
    \item The mutex will be reacquired prior to the cancellation cleanup handler getting control.  The state of the condition variable is also set to true.  The cleanup handler can safely proceed as if the mutex has been reacquired.   
\end{enumerate}
Solution: \ref{pthread_cond_wait_solution}
\end{exercise}

%TBC - still need to do this - we work on the precise-java together (this is slide 12?)

\paragraph{Programmer:} 
For the programmer, it is imperative to understand the behavior of the synchronization primitives in order to use them correctly in your program.  As we have previously discussed, there are a number of problems that can occur when the synchronization primitives are used incorrectly:
\begin{enumerate}
    \item Consider whether a thread may already hold a lock when invoking the synchronization method that obtains the lock.  This includes error recovery and exception handler paths.  In those paths where this can occur, ensure that the processing handles the lock appropriately, depending on how the synchronization primitive handles a request for a lock that is already held by the thread.
    \item Also consider how an exception is handled when an exception occurs within the scope of a lock.  The exception handler may need to explicitly release the lock or obtain the lock again.
    \item Ensure that shared data is accessed consistently using the same synchronization protocol.
    \item Consider the visibility of changes made to shared data to ensure that updates made in one thread are visible at the appropriate time to other threads.
    \item When waiting on an event while locked, consider the range of events that can occur, such as interrupts or signals, and ensure that each is handled appropriately.
\end{enumerate}

\paragraph{Tester:} 
The best approach to testing is to ensure that the tests create scenarios that result in contention on shared resources, similar to the approach described in the \hyperref[protect-general-tester]{\textit{protect-general}} pitfall.  While driving these test scenarios, include alternative events while a lock is held in order to drive error recovery paths.  For example, generate a termination signal such as $SIGTERM$ (see \href{https://www.gnu.org/software/libc/manual/html_node/Termination-Signals.html}{link}.)

\subsubsection{How POSIX handles threads [\textit{posix-pthread}] }

When writing concurrent programs, an important concern is how the system manages threads and the implications of invoking specific services when a process has multiple threads.  Although we will focus on examples in POSIX and POSIX-compliant operating systems, most operating systems have considerations such as the ones we will describe here.

One pitfall occurs with the $exec()$ function, \href{https://pubs.opengroup.org/onlinepubs/9699919799/functions/exec.html}{link}.  This function overlays the current process image, or executing program, with a new executable image.  When invoking $exec()$, all threads that are part of the process are terminated prior to the new executable image being loaded and executed.  This also means that most synchronization objects will also be deleted, with the exception of process-shared (or pshared) mutexes and condition variables.

\begin{exercise}
\label{pthread_exec}
    When calling the $exec()$ function, will the caller's cleanup handlers receive control prior to loading and executing the new executable program?
    \begin{enumerate}
        \item Yes.  The $exec()$ function causes the current program to terminate prior to loading the new executable program, and cleanup handler routines are guaranteed to receive control when the program terminates.
        \item No.  The $exec()$ function documentation clearly states that cleanup handlers will not receive control.
        \item The behavior is not defined by the POSIX standard.  It is up to the specific implementation to define its behavior.
        \item It depends on the cancelability state of the current process.  If the thread is cancelable, then cleanup handlers will receive control.  If the thread is not cancelable, then the cleanup handlers will not receive control.
    \end{enumerate}
Solution: \ref{pthread_exec_solution}
\end{exercise}

%TBC - talk to Mike Gildein, see if he wants to use the book in his teaching when we're ready.

Note an interesting point about the last exercise.  Based on just knowledge and understanding most of the alternatives that are given in the exercise seems feasible. To determine the correct alternative one need to review the $exec()$ documentation.  when reviewing the documentation the right answer becomes clear cut.  Without the documentation it is really hard to tell.  The lesson is to read the documentation and carefully understand the semantic.  Otherwise, there is little chance that the program one writes, if using concurrency, will be correct. 

It is also important to understand the different ways that an individual thread can be terminated as opposed to the ways that the process (and its individual threads) can be terminated in a multithreaded application.  When invoking the $exit()$ function from any thread, the process is terminated.  In comparison, the $pthread\_exit()$ function only terminates the current thread.  This is also true when the $main()$ function invokes $pthread\_exit()$.  In that case, the initial thread is terminated, but other threads in the process continue to execute.

Similar considerations apply when returning from a function without invoking $exit()$ or $pthread\_exit()$.  If the start function of a thread returns, the thread terminates.  However, if the $main()$ function returns, the process is terminated and not just the main thread.

When invoking functions in a multi-threaded application, it is sometimes important to understand if functions contain their own synchronization points.  For example, a common debugging technique is to add debugging statements, such as $printf$ or $fprintf$, which traces the flow through a program.  However, the standard I/O functions in ANSI C are thread safe, which means that they are synchronization points that can affect the flow of the program.  When these debugging statements are enabled, they can mask other synchronization errors because they inject additional synchronization points.  

\begin{example}
    \label{debug_example}
    Assume you have a program that creates a variable number of threads.  Each thread increments a global variable by 1.  At the end of the main routine, it prints out the global variable value, which should be equal to the number of threads. 
    Each thread follows the following pseudocode process:
    \begin{verbatim}
        local_var = global_var
        if debug_mode() print("loaded global var")
        local_var = local_var + 1
        global_var = local_var
        if debug_mode() print("updated global var")
    \end{verbatim}
    And the main routine follows the following pseudocode process:
    \begin{verbatim}
        global_var = 0
        for every thread:
            create thread
        wait for all threads to complete
        print("global variable value = %d",global_var)
    \end{verbatim}
    When testing your program, if you turn off debugging mode, it runs successfully, but if you turn it on, you get the wrong value.
    Depending on the implementation language and the operating system, there are likely one of two issues occurring:
    \begin{enumerate}
        \item The $print$ routine is a synchronization point that causes the threads to synchronize after the value of $global\_var$ has been copied.
        \item The $print$ routine injects enough of a delay that multiple threads copy the $global\_var$ value before updating it.
    \end{enumerate}
    For an example coded in C, see Appendix \ref{debug_example_code}.
\end{example}

\paragraph{Programmer:} 
For the programmer, understanding the behavior of the various functions that are being invoked is important.  As we have previously described for this pitfall, some functions may implicitly cause termination of specific threads, while other functions may result in process termination which terminates all of the threads within that process.  Additionally, some functions may have implicit synchronization processing that affects the behavior of the calling program.

A related consideration is to consider what happens if a thread or process terminates.  This may result in clean up processing being implicitly invoked, such as all threads being terminated when the main thread of the process exits.

Reviewing the documentation of the functions to be invoked is very helpful in understanding the behavior of those functions and the effects that they have on the calling threads and processes.

\paragraph{Tester:} 
For the tester, look for any invocation of external functions, especially those that have obvious implications for other threads within a process, such as the POSIX $exec()$ function.   Create tests that drive the calls to these functions.

Also create tests that test the termination of threads or processes, whether through functions like $exit()$ or $pthread_exit()$, or by the thread or process terminating, to ensure that they terminate as expected and do not affect the behavior of other threads or processes.

Finally, make sure to consider all variations of the application configuration when testing, including enabling or disabling debugging modes that are only intended for diagnostics.  These configuration options often cause the application to invoke services that have implicit synchronization that affect the behavior of the application.

\subsubsection{Concurrency thou shalt not}

To sum up concurrency programming is a hard skill to master and testing it is hard too.  We have outlined pitfalls of concurrency and attempted to emphasize that you need to get specific and exactly understand the language and synchronization primitives definitions and assumptions in order to make it fly.  In addition, there are three general rules one can apply to avoid the concurrency pitfalls we have discussed in the previous sections.  We list those rules below.

\begin{enumerate}
    \item Do not try to coordinate threads and processes using the language intuitively in a non standard manner.  For example one thread conducts a $while(x < 0)\{ sleep(1) \}$ loop to wait for another thread to set $x$ to a positive number.  This will unnecessarily consume processing resources by waking up every second to see if the value of $x$ has changed.  It will also cause extra delays between the two threads because the sleeping thread must wait for the one-second timer to expire before waking up to check if $x$ has changed.  Additionally, if $x$ may change from a negative number to a positive number and back again to a negative number within the one second that the other thread is sleeping, then it is possible that the sleeping thread never sees that $x$ became positive. Instead you want to use the standard synchronization primitives and carefully understand their semantic to achieve the desired synchronization between threads. 
    A similar situation can occur when trying to make assumptions regarding how long it takes for a thread to complete its processing.  For example, if one thread needs to wait for a second thread to complete its processing, and through experience we find that the second thread almost always completes its processing within 45 seconds, we might decide to have the first thread wait 60 seconds and then proceed assuming that the second thread must have completed its processing.  You may find that in rare cases, there are other delays that you did not account for in the second thread that cause it to take even longer than the expected 60 seconds.
    \item Do not make assumptions as to the order by which threads may obtain a lock when it becomes available.  Many locking mechanisms will allow the threads to race and obtain the lock rather than guaranteeing FIFO (first-in-first-out) order of obtaining the lock.  Some examples are the C mutex lock in the pthreads (POSIX threads) library and the Java synchronization block that do not guarantee the order by which waiting threads will obtain the lock. 
    If you have a condition that requires FIFO processing, consider whether using a lock is the best design.  For example, if you have two threads that perform almost all of their processing while holding the lock, but do very little while not holding it, consider whether you gain any benefit from having multiple threads for this processing.  In this case, a single thread to perform both pieces of processing could be easier to understand and maintain while providing similar performance.
    Another example is that of a synchronized queue that requires items on the queue to be processed in FIFO order.  If you have two threads that are retrieving items from the queue to be processed, and each thread obtains a lock before retrieving an item from the queue and does not release the lock until after the item is processed, then one of the threads is spending most of its time waiting for the other thread to release the lock.  Alternatively, if each thread releases the lock after an item has been retrieved from the queue, then the lock only ensures that items are retrieved from the queue in FIFO order, not that the processing for each item occurs in FIFO order.  Like the prior example, a single thread that processes items on the queue would be simpler to understand and maintain rather than managing multiple threads.
    \item Do not rely on priorities to cause work to be processed in a specific order.  Many times these priorities are considered to be suggestions when the scheduler dispatches work, and lower priority work may be occasionally dispatched in order to prevent starvation and guarantee that all work gets a chance to run.  In addition, using priority makes your program less portable as different schedulers may implement priorities in very different ways causing your program to behave differently when executed on top of different operating systems. 
    If your application can be run in a cloud computing environment, that paradigm exacerbates this pitfall.  In most cases, the application will be running on hardware or within a virtual machine, on an operating system that is not known to the end user.  The environment is allocated to the user on demand from a pool shared with other users, and the user of your application will normally have limited control over the environment where the application is run.  This type of environment requires the program to be as portable as possible.
\end{enumerate}

\subsubsection{Using language constructs to enforce protocol compliance [\textit{compliance-general}] }

Sometimes a language construct will help defend shared resources.   Whenever possible such constructs should be utilized to avoid concurrency issues.  That happens in two stages.  First the programmer needs to identify that a shared resource is being manipulated and will be accessed by more than one thread.   Once this is does, the programmer needs to identify language constructs that can help protect the shared resource and utilize them appropriately.  Following is a concrete example in Java using a synchronized method (see \href{https://docs.oracle.com/javase/tutorial/essential/concurrency/syncmeth.html}{link}.)

\begin{lstlisting}[language=Java]
// a thread may access i directly instead of through get()
// that synchronizes the access. 

class Compliance {
   public int i;
 
   synchronized int get(){
      return(i);
   };
};
\end{lstlisting}

\paragraph{Programmer:}
Both programmers and testers should be aware of the desired confinement (see \href{https://dzone.com/articles/java-concurrency-thread-confinement}{link}) of states in the program.  The best way out of this pitfall is to explicitly define confinement for data used by the program.  This is about visibility, so  ask yourself which threads are allowed to view the data and under what conditions.  Once confinement is well defined utilize programming constructs and access protocols to ensure it.  In the above example if everyone use the get method it would be a first step in protecting $i$.  Consider the following exercise. 

\begin{exercise}
\label{confinement_exercise}
Can the programming language can be further utilized to support the confinement of $i$?\\
Solution: \ref{confinement_solution}
\end{exercise}

Utilizing the programming primitives as in the above exercise helps enforce the desired access protocol, visibility and confinement of the data state of the program. 

\paragraph{Tester:}
For the tester, review the code for any data that has visibility to multiple threads and note where that data is referenced.  When testing, drive those functions that reference the data and look for the data to potentially be corrupted.

\subsubsection{Excessive use of the synchronized modifier [\textit{too-many-synchronized-java}]}

In Java a lock is associated with a class instance (see \href{https://docs.oracle.com/javase\%2Ftutorial\%2F/essential/concurrency/locksync.html}{link}.  When the key word $synchronized$ is added to the class method, the lock is obtained when the method is called and before it is executed and it is released once the method terminates.  

Naive understanding of multi-threaded programming may lead to simply adding the synchronized keyword to all of the methods of a class. Adding the synchronized modifier to all class methods results in 
\begin{itemize}
\item
    Capturing locks in different orders which may lead to potential deadlock.  
\item
    Unnecessary protection as some of the class instance data may not be shared by more than one thread. 
\end{itemize}

To avoid these potential pitfalls, only code segments that access shared resources should execute within the scope of obtained a lock through the synchronized keyword.  The following best practices should be followed by programmers and testers. 

\paragraph{Programmer:}
Use the $synchronized(o){}$ block to protect shared resources instead of the synchronized modifier at the method level as the method may include code segments that do not access shared resources and do not need to be synchronized. 
    
\paragraph{Tester:}
Review the code to identify excessive synchronization potentially through the comprehensive application of the synchronization modifier to method names.  Determine which methods can occur concurrently and access shared resources.  Create tests to exercise these methods concurrently by more than one thread and create contention on shared resources.   

\begin{exercise}
\label{synchronized_exercise}
Consider the following Java class used by more than one thread. 
\begin{lstlisting}[language=Java]
class file_counter_class {
    private int a;

    public void f(String fileName){
        int fd; 
        fd = open(fileName);
        a++;
        close(fd);
    }
    public void g(){
        a--;
    }
    public void h(String fileName){
        return(open(fileName));
    }

}
\end{lstlisting}
How would you update this to add synchronization blocks to protect the shared data in the minimal possible way?\\
Solution: \ref{synchronized_solution}
\end{exercise}

\ifdefined\SHOWWORK
\subsubsection{Avoiding Hangs [\textit{hangs}] }

Hangs that are more commonly called deadlocks are cases in which the software did not complete its execution and is waiting indefinitely for something to happen that will never occur.  In this pitfall we will discuss different flavors of hangs and how to avoid them by design and test for them.   

Many times when deadlocks are discussed people think about the following situation.  We are given two threads.  Thread one is attempting to obtain a lock held by thread two which in turn is attempting to obtain a lock held by thread one.  The following depict the situation in a time diagram of the two threads execution (time going from top to bottom).  

\begin{tabular}{|ccc|}
 \hline
 time            & thread one      & thread two      \\
 \hline
 \tikzmark{foo}  & \verb|lock(M1)| &                 \\
                 &                 & \verb|lock(M2)| \\
                 & \verb|lock(M2)| &                 \\
 \tikzmark{bar}  &                 & \verb|lock(M1)| \\
 \hline
\end{tabular}
\begin{tikzpicture}[overlay,remember picture]
\draw[->,black,thick] (pic cs:foo) -- (pic cs:bar);
\end{tikzpicture}

Note that in the above example thread one is holding lock $M1$ and is attempting to obtain lock $M2$. As thread 2 has already obtained lock $M2$, thread 1 will be waiting for thread 2 to release the $M2$ lock.  On the other hand thread 2 after it obtained lock $M2$ is now attempting to obtain lock $M1$ but that has already obtained by thread 1 that is waiting for thread two to release lock $M2$... DEADLOCK!

The above situation can be generalized to any cycle of threads that attempt to obtain locks.   For example, thread 1 obtained lock $M1$ and attempts to obtain lock $M2$ that thread 2 has obtained.  Thread 2 attempts to obtain lock $M3$ that thread 3 has obtained.  Thread 3 attempts to obtain lock $M1$ that thread 1 has obtained.  Again a deadlock has occurred. 

%TBC - left off here on 6/21 - slide 18/19, continuation of the same pitfall

%Slide 19:
%Hangs are caused by
%  A single thread that attempts to obtain the same lock twice (synchronization primitive dependent)
%    Recursive calls 
%thread one                                thread two                         (pthread)
%Lock(M1)                                    
%                                                 signal thread one 
%Signal handler is invoked
%Attempt to lock M1

\paragraph{Programmer:}
%slide 20:
%Define a locking policy
%  Always obtain system locks in the same order
%Carefully implement critical sections (the code region within the scope of a lock) so that
%  All operations in the critical section are none blocking
%  Error paths performs the exit protocol of the critical section 
%    Usually it means to release the critical section lock
%  either signal handlers don’t use locks, or critical sections are made non-interruptible (signals are masked)

\paragraph{Tester:}
%Review questions
%  Is there a locking policy?  Are there blocking operations (socket connect) within the scope of a lock?  Are error paths releasing locks?
%Implement tests that 
%  Force error paths while in a critical section
%  Cause the blocking of operations within a critical section
%  Kill threads that are in a critical section
%Ask for the implementation of traces that report the order in which locks are obtained and released

\fi

% Pitfall template:
%\subsubsection{Pitfall Title [\textit{pitfall-name}] }
%\paragraph{Programmer:} 
%\paragraph{Tester:} 

\section{Desk checking}

Desk checking is another review technique that is both very old and very effective.  Desk checking involves manual execution of the program, or "playing computer", and reviewing the behavior of the program as part of the walkthrough.

We will first demonstrate this technique with an example, and then describe the various roles and best cases for using this technique.

Consider the following segment of code:
\begin{verbatim}
int sum = -1;
int j = 0;
read(j); /* Ask the user for input */
for (int i=0; i<j; i++) {
  sum = sum + i;
}
\end{verbatim}
The intention of this code segment is to ask the user to input an integer j, and them sum all of the positive integers that are smaller than j.  For example, if j = 5, the intent of this code segment is to add the integers 1+2+3+4 to get a sum of 10.

We will now walk through this code segment using a value of 1 for j:
\begin{center}
\begin{tabular}{|c||c|c|c|} 
 \hline
  & \multicolumn{3}{|c|}{Resulting values} \\
 Statement & sum & i & j \\
 \hline
 \verb|int sum = -1;| & -1 &   &   \\ 
 \verb|int j = 0;| & -1 &   & 0 \\
 \verb|read(j);| & -1 &   & 1 \\
 \verb|int i=0;| & -1 & 0 & 1 \\
 \verb|(i<j)| returns true & -1 & 0 & 1 \\
 \verb|sum  = sum + i;| & -1 & 0 & 1 \\
 \verb|i++;| & -1 & 1 & 1 \\
 \verb|(i<j)| returns false & -1 & 0 & 1 \\
 \hline
\end{tabular}
\end{center}
The expected sum of all of the positive integers that are less that 1 should be 0.  However, our code segment returned a sum of -1, which is incorrect.

One might ask why it is useful to perform a desk checking exercise when you can simply execute the program and see the results.  A desk checking exercise can often show errors in the code even though the code will execute successfully for a specific set of inputs.  The act of walking through the code by hand can generate questions about how the code behaves or motivate the reviewers to suggest additional test scenarios.

\begin{exercise}
\label{deskcheck1}
    Given the following code segment:
    \begin{lstlisting}[language=C]
        i = 0; 
        if (i > 0) {
          i++; 
        }
        else {
          i--; 
        }
    \end{lstlisting}
    Perform a desk checking review and determine which of the following values for i will occur at the end of the program:
    \begin{enumerate}
        \item \verb|i = -1|
        \item \verb|i = 0|
        \item \verb|i = 1|
    \end{enumerate}
Solution: \ref{deskcheck1_solution}
\end{exercise}

The first step in a desk checking review is to generate a set of test scenarios to use for the review.  We will go into test selection in detail in a later chapter.  However, it is useful to start with simple, typical values, as well as extreme values such as boundary conditions, both on the valid side of the boundary as well as the error side of the boundary.  The set of test scenarios should be of a manageable size such that the reviewers can go through the various scenarios in a reasonable amount of time.

This brings us to the first specialized role in a desk checking review: the tester. The tester acts as one of the reviewers but has the additional responsibility of generating the test scenarios to be used in the review.  These scenarios should be prepared in advance of the review.  The same scenarios that are used in the review may also be useful when performing unit test (or other test phases) on the code.  

During the review, the review team should walk through the code several times, each time using a different test scenario.  One member of the review team should be assigned the job of the "program counter", who is the person who keeps track of what instruction is currently being executed and the values of the various variables used by the program.  This is usually the code developer, but this role can be assumed by any member of the review team.

During the review, there are a number of potential problems that should be noted.  It is useful to track statements that were never executed or decision points that were never taken.  There are two possible reasons that you might find code that was never executed.  First, this may be an indication that the test scenarios were not as thorough as they should be.  This can be solved by adding additional test scenarios. The second reason is that there may be statements that can never be executed, often referred to as "dead code."  This can be a sign that there is unnecessary code that should be removed, but more often the code developer expected these statements to be executed and therefore there is a defect that is preventing these statements from executing.  

Another potential problem to note is any value that did not have an impact on the execution.  This is often a sign that there is a missing check for that value.

Yet another potential problem is where there is an expected result that never actually occurred using the test scenarios.  This could be another sign that additional test scenarios need to be added.  It is also possible that there is some missing processing that should generate that result.

Walking through the code using these test scenarios can generate discussion about the intermediate states of the program.  This discussion can be extremely valuable as it might highlight issues with the logic of the program or assumptions regarding the potential inputs that were not valid.  Sometimes it is possible that a defect in the program can mask other defects elsewhere in the program, and when that defect is fixed the other defects are uncovered.  This highlights the possibility that if a defect is found in a section of code, it is likely that there are more defects that exist in that section of code.

During the review, discussion about the program and its intermediate states will often suggest additional test scenarios that were not part of the original list of test scenarios.  If it is practical to do so, walking through the code using these additional scenarios during the review can often find other defects.  When these additional scenarios are suggested, it is often a sign that there is a concern about a potential defect that can be exposed using these scenarios.  However, this does not necessarily indicate that the original test scenarios were incomplete or should be discarded.  The original scenarios were useful in the code walk through because they generated the discussion that led to the additional scenarios.  A good test scenario is one that helps to detect a defect that has yet to be discovered.

%TBC - how much do we want to talk about CTD?  I think it fits better in the later test selection chapter.

Desk checking exercises are very effective when used for reviewing sequential programs or "straight-line" code.  Reviewing programs that accept many input values using this technique is extremely valuable. In the next section, we will discuss a similar technique for programs that can run concurrently.

\begin{exercise}
\label{deskcheck2}
    The function \verb|reverse(L)| reverses the order of the string L.   Given that \verb|L = "abc"|  then after executing \verb|reverse(L)|, L will be set to \verb|L = "cba"|.  For the following code segment:
    \begin{lstlisting}[language=C]
L = "abc";
read(n);
counter = 0;
while(counter < n) {
    reverse(L);
    counter++;
    print(L);
}
    \end{lstlisting}
Determine the correct output if n is set to 3.
\begin{enumerate}
    \item The correct output is \verb|cbabcacba|.
    \item The correct output is \verb|cbaabccba|.
    \item The correct output is \verb|abccbaabc|.
\end{enumerate}
Solution: \ref{deskcheck2_solution}
\end{exercise}

\section{Interleaving review technique}
In the prior section, we covered desk checking as a review technique.  That technique works very well for sequential programs.  Most modern applications have many tasks or threads of work, operating in parallel.

Concurrency problems are difficult to find, since there are usually variations in the timing of when the individual threads perform their processing which can affect the behavior of the application.  It is also difficult to measure coverage of the potential interleavings between the various threads of work.  Even when a problem is uncovered during testing, it is often very difficult to debug the problem or recreate it because it often is specific to a very specific sequence of interleavings of the threads of work.

Desk checking does not take into account changes to common control structures that may be shared among many threads of work or synchronization points between multiple threads.  A more sophisticated approach to a review is needed.  The interleaving review technique is derived from the desk checking approach to review, but is focused on concurrency issues.  

Much like the desk checking approach, the interleaving review technique requires a tester to prepare a set of test scenarios to use for the review in advance.  However, the concurrent nature of the programs should be accounted for in the test scenarios.  For example, if the application can have a varying number of threads for processing work, it would be valuable to include the number of threads in the test scenarios.  The current state of the application may also be applicable to include in the test scenario.  An example of this might be that if the application has 4 threads responsible for processing requests on a queue, then it should consider what happens if all of the threads are busy processing requests when a new request comes in.  Another possible test scenario consideration is the specific interleavings between units of work.  This could mean considering what happens if thread 1 begins processing a request at a critical point in the code as thread 2 is finishing processing a request at the critical code point.

\begin{exercise}
\label{irt_1}
    Assume there are two threads.  Thread 1 has the following code:
    \begin{verbatim}
        A()
        B()
        C()
    \end{verbatim}
    Thread 2 has the following code:
    \begin{verbatim}
        D()
        E()
    \end{verbatim}
    Assume that methods A(), B(), C(), D(), and E() are all atomic operations.
    What is the number of possible interleavings that can occur between these two threads?
    \begin{enumerate}
        \item 6
        \item 8
        \item 10
        \item 12
    \end{enumerate}
Solution: \ref{irt_1_solution}
\end{exercise}

Programs that benefit from an interleaving review technique walkthrough tend to have a very large test scenario space due the concurrent nature of these programs.  The tester will often have to make choices as to the scenarios that provide the most value for the limited amount of time that the reviewers can provide.  However, many of the same considerations for generating test selection in a desk checking review also apply to the interleaving review technique.

Like the desk checking review, there is a member of the review team that should be assigned the role of "program counter" to keep track of the current state of the system and the current instruction location of each unit of work.  An additional role that needs to be fulfilled is what we refer to as the "devil's advocate".  This person should decide the timing of events, such as when to switch from one unit of work to another.  We will provide guidelines to help make these choices later in this section.  This person's goal is to choose timing that is most likely to uncover an issue.

During the review, the review team will walk through the code using different test scenarios, much like a desk checking review.  When walking through the scenario, the reviewers will follow a single thread of work until the devil's advocate makes the decision when to suspend that thread and switch to a different thread.  The reviewers will then follow the new thread until the devil's advocate makes the decision to suspend that thread and switch back to the original thread, or switch to yet another thread.

If the team has the ability to use a simulator or debugger to execute the application as part of the review, then this can be used to help facilitate the review by showing the state of the threads as part of the review.  However, this is not a requirement for this type of review.

The review team should look for many of the same potential problems that they would look for in a desk checking review, such as statements that were not executed or values that did not have impact on execution.  

Also, similar to a desk checking review, discussion about the programs and their intermediate states will often suggest additional test scenarios or additional timing scenarios that were not part of the original list of scenarios.  If it is practical to do so, walking through these scenarios can find defects.  

\begin{example}
    Consider the following simple code snippet:
    \begin{lstlisting}[language=C]
boolean chosen;            // Global variable shared by all processes
                           // used for process coordination 
                           // This is initialized to false
boolean iAmLeader = false; // local variable indicates the current
                           // process leader status
if (chosen == false) {
   lock();
      chosen = true;
      iAmLeader = true;
   unlock();
}
    \end{lstlisting}
    This segment of code can run under multiple processes, and is meant to choose one process to be a leader.  Assume that there are two processes running this segment of code.  
    We will now walk through this code segment:
\begin{center}
\small
\begin{tabular}{|c|c||c|c|c|} 
 \hline
  \multicolumn{2}{|c||}{Statements} & \multicolumn{3}{|c|}{Resulting values} \\
 Process 1 & Process 2 & chosen & Process 1 & Process 2\\
 & & & iAmLeader & iAmLeader \\
 \hline\hline
 Begin process execution & & false & false & false  \\  \hline
 \verb|if (chosen == false)| & & false & false & false \\ returns true & & & & \\ \hline
 Suspend process execution & Begin process execution & & & \\  \hline
 & \verb|if (chosen == false)| & false & false & false \\
 & returns true & & & \\ \hline
 & \verb|lock();| & false & false & false \\ \hline
 & \verb|chosen = true;| & true & false & false \\ \hline
 & \verb|iAmLeader = true;| & true & false & true \\ \hline
 & \verb|unlock();| & true & false & true \\ \hline
 Continue process execution & Suspend process execution & & & \\ \hline
 \verb|lock();| & & true & false & true \\ \hline
 \verb|chosen = true;| & & true & false & true \\ \hline
 \verb|iAmLeader = true;| & & true & true & true \\ \hline
 \verb|unlock();| & & true & true & true \\ \hline
\end{tabular}
\end{center}
As you can see, by performing this interleaving, both processes end up with their iAmLeader value set to true, which is not what was intended.
\end{example}

There are some common guidelines for the timing of processes:
\begin{itemize}
    \item Consider contention on shared resources.  Any scenario that increases contention is likely to expose issues.
    \item Delaying obtaining of locks or other synchronization methods so that synchronization is not obtained in the same order each time.
    \item Forcing of error paths, particularly in critical sections of code.  This can often cause signals or interrupts to occur, resulting in blocked operations to become unblocked.
    \item Be sure to cover all possible scenarios of waiting on an event.  For example, consider not just waiting for an event followed by the event occurring, but also if the event occurs before waiting for it (so that there is no need to wait.)
    \item If appropriate, avoid assumptions specific to operating system scheduling or hardware.  This can include unnaturally long delays in dispatching of work or memory changes in other threads not being immediately visible by the current thread.
\end{itemize}

A starting list of points to consider switching processes in a review:
\begin{itemize}
    \item When updating a common field
    \item When setting or resetting a "footprint" or other variable that indicates the current state of the current thread
    \item Before obtaining synchronization or after releasing synchronization
    \item When invoking an external interface, including system services.  These are often "black boxes" from the point of the code that is being reviewed.
\end{itemize}

\section{When to use certain techniques}

We have previously described several review techniques and how they can be used.  We will now review when they should be used.

Paraphrasing is a good "general-purpose" technique.  There are other techniques tailored to specific situations, but paraphrasing can be used in almost all situations.  Reviews will usually use paraphrasing as a starting point, especially in the absence of any specific concerns or if concerns are limited to specific parts of the material to be reviewed.

If appropriate checklists exist, reviewing code based on those checklists is also a good technique and can be combined with paraphrasing or other techniques.

Based on specific concerns about the material to be reviewed, it might make sense to use other techniques instead of, or in conjunction with, paraphrasing.

When reviewing new interfaces, or use of a new interface, an obligation review makes sense to ensure that all of the obligations of the caller and the routine being called are met.  A desk checking review can also make sense if the interface has a large number of inputs.

Error recovery, whether it be synchronous checking of processing, like return code handling, or asynchronous checking like signal handler processing, also lend themselves to an obligation review.  Desk checking can also make sense, especially for synchronous checking.  

When creating new routines, especially those with a large number of inputs, a desk checking review is often a good choice.

When concurrency or synchronization is involved, an interleaving review is usually very effective.

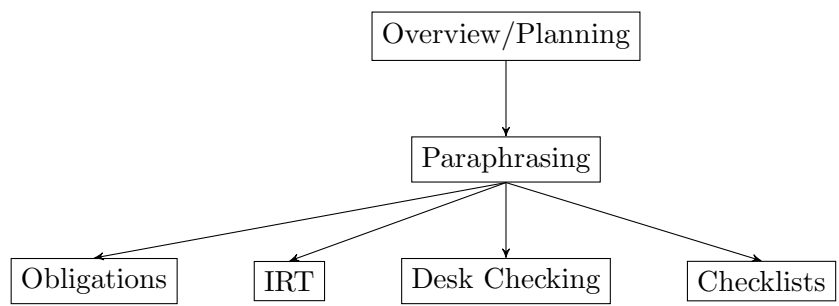
\begin{figure}
    \begin{tikzpicture} [every node/.style={rectangle,draw=black,align=center}] 
        \node(overview){Overview/Planning}; 
        \node(paraphrasing)[below= of overview]{Paraphrasing}; 
        \node(deskcheck)[below= of paraphrasing]{Desk Checking};
        \node(irt)[left= of deskcheck]{IRT};
        \node(obligations)[left= of irt]{Obligations};
        \node(checklists)[right= of deskcheck]{Checklists};
        \draw[->] (overview.south) -- (paraphrasing.north);
        \draw[->] (paraphrasing.south) -- (obligations.north);
        \draw[->] (paraphrasing.south) -- (checklists.north);
        \draw[->] (paraphrasing.south) -- (deskcheck.north);
        \draw[->] (paraphrasing.south) -- (irt.north);
    \end{tikzpicture}
    \centering
    \caption{Review Process and Technique flow}
\end{figure}

%TBC - Scott, pick up writing new material here.

\chapter{Time management and the review process}

In the previous chapter, we have discussed how to prepare for a code review, as well as various techniques that can be used in the code review process.  We will now discuss time management.

Time management is an important part of the review process.  There are a few aspects to time management.  The most obvious is to manage the time spent reviewing itself.  If you are going to have a formal meeting, a good rule is to limit the time in the review to two hours or less.  One hour is ideal.  Reviewers have a limited attention span.  A review that is more than two hours will many times result in a loss of reviewer effectiveness and productivity.  If more review time is needed, it is a good idea to break the review up into smaller reviews, spread over multiple days.

The other aspect of time management is planning ahead!  Always assume that a review is going to have some number of followup questions, concerns, or generate some rework. 
If it does not there are other things wrong with the review process as it is not revealing any problems. 
Rework and followups need to be factored into project deadlines.  A review has limited value if you have already decided to deliver your code without changes.  

\section{Preparation}

To optimize the use of reviewers time and facilitate the effectiveness of the review process, a good review should have a preparation phase.  Preparation includes assembling the artifacts to be reviewed.  In addition to the code, it may be useful to include design documents or other project overview information, traces, screenshots, test plans, or documentation.  Your reviewers are helping you by reviewing your code, so you want to make it as easy as possible to review your code.

You should also consider the time you expect the reviewers to spend in the review and the amount of time they can afford.  That implies that you should plan the review accordingly and focus the review on your highest concerns.  By doing so you can make the best of the time that reviewers can realistically allocate for the review process. 

For certain review techniques, it can be valuable to provide a set of test conditions to be used as part of the review process.  Preparing these test conditions can help the reviewers focus on certain conditions and make it easier for them to understand the flow through the code to be reviewed.  These test conditions should also be prepared in advance.  We will cover test selection in detail in a subsequent chapter.

Related to test conditions, a common question or consideration is how much testing should be done prior to holding a review.  A piece of code that has undergone little to no testing is more likely to contain defects than a piece of code that as undergone significant testing.  To get the most value out of your review, you do not want your reviewers to spend their time finding easy problems that could have been found by running some simple tests.  However, if a major concern is raised during the review, it may require changes that necessitate repeating some or all of the testing that has already been done.  Therefore, some balance is necessary.  The authors suggest that any testing that can be performed easily, such as mainline code paths, be done prior to the review.  Alternatively, testing that is complicated, time consuming, or otherwise hard to perform, such as error recovery testing, is often done after the review.

It is also necessary to consider the goals of the review as part of planning the review.  The primary goal of a review is to find defects.  There can also be some very valuable secondary goals.  One of these other goals is to provide education.  With a large code base, the team members may have very few chances to gain direct experience in all of the pieces of code that the team owns.  In addition, a code review is a great opportunity to gain some understanding of code that is rarely updated.  During a review, it is important to take the time for education.  Plan to need extra time for educational discussion in the code review.  Unlike our recommendation for issues, where we suggest spending about a minute on an issue, our suggestion is that 5 minutes is reasonable for educational discussion when an issue is raised that indicates lack of knowledge. 

Another secondary goal can be to identify code hygiene issues.  This can include identifying areas of code that would benefit from refactoring.  Refactoring is usually defined as internal restructuring of a piece of code, while not changing its external behavior.  There are many reasons to refactor code, including improving performance and making it easier to understand.  Other improvements related to code hygiene can include improvement of code comments or other internal documentation, as well as removal of "dead code" or code that is no longer used.  Of course, changes for code hygiene purposes deserve to be reviewed as well.

%TBC - do you want to group the paragraphs that discuss the meeting?

\section{During the review}

During the review, issues will be raised by the reviewers.  Issues are not limited to problems or defects.  They can also include questions that need to be answered, concerns requiring further investigation, and comments that do not result in finding a defect.

It is best to raise issues without spending much time on resolving them.  If you only have an hour for a review, and you spend half of that hour trying to resolve a concern on the first line of code that was changed, you likely will not finish the review in the allotted time.  For all but the simplest of issues, it is best to log the issue during the review and resolve it after the review.  A good rule of thumb is to spend no more than a minute or so on an issue before moving on.  There is an exception to this: if the issue raised is so significant that it is not worth continuing with the review, such as an issue that will cause the code being reviewed to be completely rewritten, then it could be worth spending the time discussing a new solution rather than continuing with the review.  In such as case one may also stop the meeting especially if the stakeholders needed to discuss the new solution are not present in the review. 

\section{After the review}

It is the responsibility of the code developer to follow up on all issues after the review.  All issues should be logged in a common repository that is accessible to all of the reviewers, and the developer should also log their response to each issue.  This makes it simple for all of the reviewers to see the responses and ensure that their issues were resolved to their satisfaction.

%TBC - maybe move to chapter 3

After the review, it is necessary to evaluate the effectiveness of the review.  We do this to consider whether our review process was productive and whether changing our approach to the review might provide better results.  This is not as easy as it might sound.  If a review uncovers a lot of issues, especially major issues, then clearly the review was effective!  However, an absence of issues uncovered in the review could show that the reviewers did not do an adequate job in the review, or that there simply were not many issues to find.  In this case our job is more difficult.  

We should consider whether the reviewers were active and engaged in the review.  If the reviewers asked questions and there was a lively discussion, it is likely that the reviewers performed a thoughtful analysis of the code to be reviewed.  Alternatively, if the reviewers seemed confused or bored, then the review was probably not as effective as it could be.  

Also, consider the code to be reviewed.  A simple, sequential piece of code with only a few inputs is less likely to have defects than a complex piece of code with many inputs.  If your review seemed like it should have found more issues than it did, then consider a different approach to the review.  For example if you are concerned about concurrency and concurrency was not focused on in the review process you may want to consider a different review approach that focuses on concurrency.   

Another consideration for evaluating effectiveness is how much testing has been performed on the code that was reviewed.  We previously discussed testing as part of the review planning process.  A piece of code that has had little to no testing is more likely to contain defects than a piece of code that has had significant testing.  The number of issues found should reflect the amount of testing that has already been performed.

%TBC: \chapter{Test selection techniques}
%\begin{itemize}
%    \item Test selection - simple, typical, extremes/boundary conditions, etc.
%    \item examples
%    \item Goal - exhaustive testing not possible, so we're trying to cover the possible inputs/outputs and control/data flow within the program
%\end{itemize}

\chapter{Modern review processes}
As has been discussed elsewhere in this book, code reviews have been around almost as long as code has been around.  Many of the techniques that we have described do not require a significant technology investment.  Code reviews traditionally would take place by having all of the reviewers gather together at a common location, with the code to be reviewed available, and possibly a whiteboard or some paper for taking notes or tracking comments.

Although it is still possible to conduct effective code reviews in that way, it is not always possible or practical to do so.  Development teams no longer have to report to an office to do their work; in many cases the team can be scattered across the world.  In this chapter, we will cover various topics related to code reviews code in the modern era, including ways that technology can be leveraged to improve the review process.

\section{Remote meetings}
Although it is usually ideal for the reviewers to be in the same room when holding a review, there are a number of reasons for which a remotely held review may be preferred, if not required.

Remote work, especially for software engineering teams, has become more common in recent years.  Even for those organizations that are co-located in an office, many times they take a hybrid approach where the team members split their time between working from home and working from an office location.

There are also some ease-of-use considerations that can make a remote meeting valuable even when the reviewers are in the same location.  Most laptop computer screens are relatively small, compared to an external monitor.  When reviewing code online, more screen space can help make the review easier.  

Holding a remote review meeting is usually more effective than holding an offline review, where all of the reviewers will perform their review independently.  Holding a remote review meeting promotes discussion and collaboration among the reviewers.  One reviewer might ask a question about the processing in a segment of code that does not directly uncover an issue, but the discussion that results from that question may lead to additional questions or investigation.  Also, code reviews and the discussion that occurs in the review meeting can be a great source of education for the reviewers.

When holding a remote meeting for a code review, there are a number of tools that need to be available to all of the reviewers.

First, a remote meeting or conferencing tool is a requirement.  The choice of tools is usually dictated by the organization or company, but there are number of features that a good remote conferencing tool should include:
\begin{itemize}
    \item Ability to share a user's screen, so that all of the reviewers can follow the same review materials.
    \item Ability to record the meeting.  A recording can be useful when following up on comments or concerns raised at the meeting if those concerns are not clear.
    \item A virtual whiteboard or drawing tool, unless a separate whiteboarding tool is available.
    \item A chat feature, unless a separate chat tool is available.
    \item Is reliable and has good quality video/audio.
\end{itemize}

A good file comparison tool is also a requirement.  There are many different tools available.  One list of tools can be found at \href{https://en.wikipedia.org/wiki/Comparison_of_file_comparison_tools}{Wikipedia}.  A good file comparison tool should support the following:
\begin{itemize}
    \item Ability to choose between horizontal (side by side) comparison and vertical (above and below) comparisons.
    \item Customization of comparison rules, such as the ability to ignore whitespace or character case, based on the syntax rules of the programming language that the code is implemented in.
    \item The ability to "export" and "import" those comparison rules, so that all of the reviewers can use a common set of rules.
    \item Rudimentary file editing support.  It does not have to have robust file editing capabilities, but it is often useful to be able to make notes on a particular line for later investigation or followup.
\end{itemize}
There are some other features in a file comparison tool that may be useful, or even necessary, in some circumstances:
\begin{itemize}
    \item For organizations where the reviewers may be using different operating systems, cross-platform availability of a review tool is important.
    \item 3-way file comparison.  This is useful when multiple releases or versions of the code are supported and must be updated, allowing the reviewers to easily compare the changes within one release as well as across releases.
    \item Syntax highlighting.
\end{itemize}
Although it is also recommended that all of the reviewers to use the same file comparison tool, it is not a requirement.  Use of a particular tool is often a personal preference.

Some other tools that might be considered:
\begin{itemize}
    \item A a virtual whiteboarding or drawing tool that can be used for drawing flow diagrams or other visual representations of code or data flow.
    \item A file sharing or file collaboration tool that allows the reviewers to make notes in a common file in real time.  This allows each reviewer to scribe their own notes or questions for later followup by the code developer, or to make comments of style considerations or other minor issues that do not require discussion in the review.  For some review techniques, such as a desk checking exercise, a shared file can also be used for tracking the state of the system during the review.  
    \item A group chat or messaging tool may also be useful in addition to, or instead of, a file sharing or file collaboration tool.
\end{itemize}

When holding a remote review, additional consideration must be given to review preparation.  When all of the reviewers are in the same room, it is easy to quickly draw a flow diagram or other information on a whiteboard.  In a remote meeting environment, this usually would need to be prepared in advance so that they can be shared with the other reviewers.  It is not always necessary to spend a lot of time drawing a complex diagram in a whiteboarding or drawing tool.  A simple drawing on a piece of paper, captured with a scanner or even a camera phone, can be quickly done and shared prior to the review meeting.

During the review, one of the reviewers should display the code being reviewed on their screen and share their screen as the code is being reviewed.  This allows the other reviewers to follow the shared screen if they choose to do so.  The other reviewers may choose to follow along in their own comparison tool instead, but can refer back to the shared screen if they lose their place in the review materials.  The person sharing the screen will usually be the person reading, who will usually be the code developer or the meeting moderator.

It is also important that the person reading makes sure that the location in the code is communicated to the other reviewers.  This is often done by announcing the line number within the file that is being reviewed, and when switching to a different file, announcing the file name being reviewed.

For review techniques that require one of the reviewers to keep track of the state information, such as a desk checking review, a shared file in a file collaboration tool can be used to track that information.  The person reading the code can continue to display the code on their screen and share it, and all of the reviewers can switch between state information file and the screen share as needed.

A technique like the interleaving review technique, where the reviewers may be switching between multiple code files or multiple locations within a code file, can be especially difficult to use in a remote review.  One technique for solving this issue is to assign different reviewers a different thread in the code, and have them take turns sharing the screen as their thread begins executing.  

\ifdefined\SHOWWORK
\section{Offline reviews}
%TBC - need section here
\fi

\printbibliography

\appendix
\chapter{Solutions}
\label{chapter:solution_appendix}
Following are solutions to selected exercises. 

\begin{solution}
\label{openandwrite_solution}
Solution of exercise \ref{openandwrite}.  The correct answer is the third answer: "Indeed we would like to check if the buffer memory is allocated or not, but clearly we are not checking if the open operation was successful."
\end{solution}

\begin{solution}
\label{compare_solution}
Solution of exercise \ref{compare}.  The first and the last answers are correct. 
\end{solution}

\begin{solution}
\label{paraphrasing_solution}
Solution to exercise \ref{paraphrasing}.  The correct answer is the second answer.  You need to explain what the code is doing and not be literal when paraphrasing. 
\end{solution}

\begin{solution}
\label{obligation_signals_solution}
Solution to exercise \ref{obligation_signals}.  The second answer is the correct answer.  The signal handler may be responsible for handling errors found in the mainline processing.  The control flow includes the signal handler, so it should be part of the scope for an obligation review, therefore the third answer is not correct.  The type of signal may influence the flow through the signal handler, causing certain obligations to be handled for some signals and not others.  Therefore the first answer, although true, is not as complete as the second answer.
\end{solution}

\begin{solution}
\label{obligation_solution}
Solution to exercise \ref{obligation}.  The correct answer is the third answer.  In this case, thread 3 obtains resource R, causing the use count i to be incremented from 0 to 1.  Then, since i is not greater than 1 (as it is equal to 1), thread 3 does not release resource R.  Thread 2 then obtains resource R, incrementing the use count i from 1 to 2, and releases resource R, which decrements the use count from 2 down to 1.  Thread 1 then performs similar processing to thread 2.
\end{solution}

\begin{solution}
\label{obligation_f_solution}
Solution to exercise \ref{obligation_f}.  The correct answer is the third answer.  The first answer is partially correct, in that it includes one of the necessary obligations, but it is not enough to only check that one obligation.  For the other two answers, the behavior is not defined when \verb|serverIsRunning()| takes an exception.  Although it could be reasonable behavior for \verb|i| to not be incremented when \verb|serverIsRunning()| takes an exception, this should be explicitly specified, so the third answer is a better answer than the second answer.
\end{solution}

\begin{solution}
\label{pitfalls_solution}
Solution to exercise \ref{pitfalls}.  Both 2 and 3 are correct answers.
The first answer, an out of boundary array access, is a programming pitfall, but does not directly map to the correct releasing of resources.
The last answer, error paths always lead to the termination of the process, is not necessarily true. 
\end{solution}

\begin{solution}
\label{checklist_solution}
Solution to exercise \ref{checklist}.  The second answer is correct.  A good checklist makes a trade between usability and completeness.  A checklist that is complete and describes all known pitfalls will often be very large.  This can be counterproductive, since it can be hard to review for all of the items in the checklist, and will often be ignored if it grows too large.  Although in some cases it may be useful to review code with a focus on enforcing standards, a checklist review is better suited to reviewing for certain bug patterns instead.
\end{solution}

\begin{solution}
\label{pthread_cond_wait_solution}
Solution to exercise \ref{pthread_cond_wait}.  According to the definition of \\ 
$pthread\_cond\_wait()$ at 
\href{https://pubs.opengroup.org/onlinepubs/7908799/xsh/pthread_cond_wait.html}{https://pubs.opengroup.org}, the correct answer is the third answer.  It does state that if the thread is set to first deferred cancellation, the behavior is that the mutex is reacquired prior to the cancellation handler receiving control.  For other cancelability states, it makes no statement about the state of the mutex, and though it is likely that the thread does not hold the mutex lock, this is not guaranteed.  
It also makes no statement about the state of the condition variable.
\end{solution}

\begin{solution}
\label{pthread_exec_solution}
Solution to exercise \ref{pthread_exec}.  According to the definition of $exec()$ at\\
\href{https://pubs.opengroup.org/onlinepubs/9699919799/functions/exec.html}{https://pubs.opengroup.org}, the correct answer is the second answer.  It states that no cleanup handlers shall be called.
\end{solution}

\begin{solution}
\label{confinement_solution}
Solution to exercise \ref{confinement_exercise}.  Yes.  In this example, changing the definition of $i$ from $public$ to $private$ will prevent users of the class from referencing $i$ directly, forcing the caller to use the $get()$ method to reference the value of $i$.
\end{solution}

\begin{solution}
\label{synchronized_solution}
Solution to exercise \ref{synchronized_exercise}.  
The only processing that needs to be synchronized is the processing that updates $a$.  The $f$ function would be updated as follows:
\begin{lstlisting}[language=C]
    public void f(String fileName){
        int fd; 
        fd = open(fileName);
        synchronized(this) {
            a++;
        }
        close(fd);
    }
\end{lstlisting}
This synchronizes the processing that increments $a$ without unnecessarily synchronizing the calls to $open$ and $close$.
The $g$ function can be updated similarly:
\begin{lstlisting}[language=C]
    public void g(){
        synchronized(this) {
            a--;
        }
    }
\end{lstlisting}
Alternatively, it could also be updated to be a synchronized method:
\begin{lstlisting}[language=C]
    public synchronized void g(){
        a--;
    }
\end{lstlisting}
The $h$ function does not require updates.
\end{solution}

\begin{solution}
\label{deskcheck1_solution}
Solution to exercise \ref{deskcheck1}.  The correct answer is the first answer.  In this case, i is set to 0, and the if statement checks if i is greater than 0.  It is not,  so the else clause is taken, which decrements i from 0 to -1.
\end{solution}

\begin{solution}
\label{deskcheck2_solution}
Solution to exercise \ref{deskcheck2}.  The correct answer is the second answer.  In this case, the while loop is executed 3 times.  The first time, L is reversed so that it contains \verb|cba| which is printed.  The second time, L is reversed again so that it contains \verb|abc| which is printed. The third time, L is reversed one final time so that it contains \verb|cba| which is printed.
\end{solution}

\begin{solution}
\label{irt_1_solution}
Solution to exercise \ref{irt_1}.  The following interleavings are possible:
\begin{enumerate}
    \item \verb|A()-B()-C()-D()-E()|
    \item \verb|A()-B()-D()-C()-E()|
    \item \verb|A()-B()-D()-E()-C()|
    \item \verb|A()-D()-B()-C()-E()|
    \item \verb|A()-D()-B()-E()-C()|
    \item \verb|A()-D()-E()-B()-C()|
    \item \verb|D()-A()-B()-C()-E()|
    \item \verb|D()-A()-B()-E()-C()|
    \item \verb|D()-A()-E()-B()-C()|
    \item \verb|D()-E()-A()-B()-C()|
\end{enumerate}
The correct answer is the third answer, 10.
\end{solution}

\chapter{Coded examples}
\label{chapter:examples_appendix}
The following are selected coded examples.

\section{Cleanup handler}
\label{pthread_cond_wait_code}
The following example is a fully coded example demonstrating the $pthread\_cond\_wait()$ C function in UNIX.  This function is discussed in exercise \ref{pthread_cond_wait}.
\begin{lstlisting}[language=C]
// C program to test cleanup handler functions
#include <pthread.h>
#include <stdio.h>
#include <unistd.h>

// Declaration of mutex
pthread_mutex_t my_mutex;
// Declaration of thread condition variable
pthread_cond_t condvar;

// Definition of condition (predicate)
int condition_case = 0;

void handler1(void *arg)
{
	printf("cleanup handler: entered\n");
	if (pthread_mutex_trylock(&my_mutex) == 0) {
		printf("cleanup handler: mutex lock successfully obtained\n");
	} else {
		printf("cleanup handler: mutex lock not obtained\n");
		printf("cleanup handler: unlocking mutex lock\n");
		pthread_mutex_unlock(&my_mutex);
	}
	printf("cleanup handler: exiting\n");
}

// Thread function
void* thread1()
{
	int oldstate, oldtype;

	// Set up cleanup handler
	pthread_cleanup_push(handler1, NULL);

	// Set cancelability state.  The following states are legal:
	// - PTHREAD_CANCEL_ENABLE
	// - PTHREAD_CANCEL_DISABLE
	pthread_setcancelstate(PTHREAD_CANCEL_ENABLE,&oldstate);

	// Set cancelability type.  The following types are legal:
	// - PTHREAD_CANCEL_DEFERRED
	// - PTHREAD_CANCEL_ASYNCHRONOUS
	pthread_setcanceltype(PTHREAD_CANCEL_DEFERRED,&oldtype);

	// acquire a lock
	printf("thread1: acquiring mutex lock\n");
	pthread_mutex_lock(&my_mutex);

	// wait on condition variable
	while (!condition_case) {
		printf("thread1: waiting on condition variable\n");
		pthread_cond_wait(&condvar, &my_mutex);
		printf("thread1: awakened\n");
	}

	// release the lock
	printf("thread1: unlocking mutex lock\n");
	pthread_mutex_unlock(&my_mutex);

	// clean up and return to caller
	printf("thread1: returning\n");
	pthread_cleanup_pop(0);

	return NULL;
}

// Driver code
int main()
{
	pthread_t tid1;

	printf("main: initializing\n");

	// Initialize the condition variable and mutex
	pthread_cond_init(&condvar,NULL);
	pthread_mutex_init(&my_mutex,NULL);

	// Create thread 1
	printf("main: creating thread 1\n");
	pthread_create(&tid1, NULL, thread1, NULL);

	// Sleep for 5 sec so that thread1 gets a chance to run
	printf("main: sleeping\n");
	sleep(5);

	printf("main: cancelling thread 1\n");
	pthread_cancel(tid1);

	// wait for the completion of thread 1
	printf("main: waiting for thread 1 completion\n");
	pthread_join(tid1, NULL);

	printf("main: exiting\n");
	return 0;
}

\end{lstlisting}

The example output from running the program is as follows:
\begin{verbatim}
main: initializing
main: creating thread 1
main: sleeping
thread1: acquiring mutex lock
thread1: waiting on condition variable
main: cancelling thread 1
main: waiting for thread 1 completion
cleanup handler: entered
cleanup handler: mutex lock not obtained
cleanup handler: unlocking mutex lock
cleanup handler: exiting
main: exiting
\end{verbatim}

\section{Debug processing}
\label{debug_example_code}
The following example is a fully coded example demonstrating the $pthread\_cond\_wait()$ C function in UNIX.  This function is discussed in example \ref{debug_example}.
\begin{lstlisting}[language=C]
// C program to demonstrate debugging behavior change
#include <pthread.h>
#include <stdio.h>
#include <unistd.h>
#include <stdlib.h>

#define THREAD_COUNT 100
#define DEBUG_MODE

int global_x = 0;

// Thread function 1
void* thread1(void *arg)
{
	int *threadnumptr = (int *)arg;
    int threadnum __attribute__ ((unused)) = *threadnumptr;
	int local_x;

	#ifdef DEBUG_MODE
    printf("thread%d: entered\n",threadnum);
	#endif

	// process items
	local_x = global_x;
	#ifdef DEBUG_MODE
	printf("thread%d: loaded global value %d\n",threadnum,local_x);
	#endif
	local_x = local_x+1;
	global_x = local_x;
	#ifdef DEBUG_MODE
	printf("thread%d: increment complete\n",threadnum);
	#endif

	#ifdef DEBUG_MODE
	printf("thread%d: returning\n",threadnum);
	#endif

	return NULL;
}

// Driver code
int main()
{
	pthread_t tid[THREAD_COUNT];
	int parm[THREAD_COUNT];

    printf("main: initializing\n");

	// Create threads
	for (int thread = 0; thread < THREAD_COUNT; thread++) {
		parm[thread] = thread + 1;
		#ifdef DEBUG_MODE
	    printf("main: creating thread %d\n",thread+1);
		#endif
		pthread_create(&tid[thread], NULL, thread1, &parm[thread]);
	}

	// sleep for 5 sec so that the threads get a chance to run
	sleep(5);

	// wait for the threads to complete
	#ifdef DEBUG_MODE
    printf("main: waiting for thread completion\n");
	#endif
	for (int thread = 0; thread < THREAD_COUNT; thread++) {
		pthread_join(tid[thread], NULL);
	}

    printf("main: total count is %d\n",global_x);

    printf("main: exiting\n");
	return 0;
}
\end{lstlisting}

\end{document}